**Robust Chemiresistive Behavior in Conductive Polymer/MOF Composites**

*Heejung Roh, Dong-Ha Kim, Yeongsu Cho, Young-Moo Jo, Jesús A. del Alamo, Heather J. Kulik,* Mircea Dincă,* Aristide Gumyusenge**

H. R., Dr. A. G.
Massachusetts Institute of Technology, Department of Materials Science & Engineering, 77 Massachusetts Ave, Cambridge, MA, 02139, USA
Email: aristide@mit.edu

H. R., Dr. D.-H. K., Dr. Y.-M. J., Dr. M. D.
Massachusetts Institute of Technology, Department of Chemistry, 77 Massachusetts Ave, Cambridge, MA, 02139, USA
Email: mdinca@mit.edu

Dr. Y. C., Dr. H. J. K.
Massachusetts Institute of Technology, Department of Chemical Engineering, 77 Massachusetts Ave, Cambridge, MA, 02139, USA
Massachusetts Institute of Technology, Department of Chemistry, 77 Massachusetts Ave, Cambridge, MA, 02139, USA
Email: hjkulik@mit.edu

Dr. J. A. A.
Massachusetts Institute of Technology, Department of Electrical Engineering and Computer Science, 77 Massachusetts Ave., Cambridge, MA, 02139, USA
MIT-IBM Watson AI Lab, 75 Binney St., Cambridge, MA, 02139, USA

**Abstract**

Metal-organic frameworks (MOFs) are promising materials for gas sensing but are often limited to single-use detection. We demonstrate a hybridization strategy synergistically deploying conductive MOFs (*c*MOFs) and conductive polymers (*c*Ps) as two complementary mixed ionic-electronic conductors in high-performing stand-alone chemiresistors. Our work presents significant improvement in i) sensor recovery kinetics, ii) cycling stability, and iii) dynamic range at room temperature. We demonstrate the effect of hybridization across well-studied *c*MOFs based on 2,3,6,7,10,11-hexahydroxytriphenylene (HHTP) and 2,3,6,7,10,11-hexaiminotripphenylene (HITP) ligands with varied metal nodes (Co, Cu, Ni). We conduct a comprehensive mechanistic study to relate energy band alignments at the heterojunctions between the MOFs and the polymer with sensing thermodynamics and binding kinetics. Our findings reveal that hole enrichment of the *c*MOF component upon hybridization leads to selective enhancement in desorption kinetics, enabling significantly improved sensor recovery at room temperature, and thus long-term response retention. This mechanism was further supported by density functional theory calculations on sorbate-analyte interactions. We also find that alloying *c*Ps and *c*MOFs enables facile thin film co-processing and device integration, potentially unlocking the use of these hybrid conductors in diverse electronic applications.

## 1. Introduction

Metal organic frameworks (MOFs) are attractive for catalysis, energy storage, chemical capture, as well as sensing owing to their inherent porosity, high surface area, and high molecular absorptivity.[1–5] Particularly for MOF-based sensing, the chemical versatility of metal nodes and organic ligands renders MOFs attractive for molecularly tuning both sensitivity and selectivity towards a wide range of analytes.[6,7] Pore size control also provides an additional knob for size-exclusion based sensing.[8–10] Electrically conductive MOFs (*c*MOFs) are of particular interest in chemiresistive sensors leveraging their conductance modulation within the framework upon host-guest interaction to identify and quantify the guest molecules.[11,12] However, deploying *c*MOFs in sensors still faces technological challenges: i) MOFs are commonly synthesized as powders and their integration into electronic devices is challenging. Typically, MOF powders are pressed into pellets or suspended as pastes to form active layers, leading to poor performance reproducibility and loss in inherent properties at the expense of additive loadings.[13–15] Layer-by-layer liquid epitaxy and surface-supported MOF growth, though yet to be demonstrated for a wide library of MOF structures, have emerged as promising alternative processing strategies for applications requiring high quality thin films.[16–18] ii) MOF-based chemical sensors are often dosimetric due to limited reversibility which hinders practical use.[19,20]

Particular to gas sensing, *c*MOF-based detection of gases typically involves a combination of physical adsorption and chemical interactions.[5] For instance, ligand designs to form electron rich coordination sites favorable for attracting ubiquitous polar gas molecules through Van der Waals interactions have been demonstrated.[5] In addition, transition-metal nodes in these *c*MOFs primarily drive the majority of chemiresistive sensing owing to strong Lewis acid-base reactions between the metal nodes and analytes, especially polar gas molecules.[19,21,22] Consequently, 2D-conjugated ligands, namely 2,3,6,7,10,11-hexahydroxytriphenylene (HHTP) and 2,3,6,7,10,11-hexaiminotripphenylene (HITP), in combination with common nodes (Cu, Ni, Co) have been widely studied in chemiresistive gas sensors owing to their excellent and tunable electrical conductivity (which enables detection based on resistance change), as well as their facile synthesizability.[18,23–25] In that regard, gases such as $NH_3$, $H_2S$, and other volatile organic compounds have been reliably detected using *c*MOF-based sensors.[24]

$NO_2$ is one of the commonly emitted toxic gases that remains challenging to detect, even using conductive MOFs, especially at room temperature. Though *c*MOFs have shown promising performance for $NO_2$ detection,[26] irreversible sensing remains a major challenge due to the formation of stable coordination complexes, a trait that enables high sensitivity. Reversible detection of $NO_2$ at room temperature using MOFs becomes an inherent challenge due to strong binding behaviors, arising from $NO_2$'s tendency to extract electrons from metal nodes (e.g., $Cu^{1+}$) and form coordination complexes (e.g., (1) N-nitro, (2) O-nitrito, or (3) O, O′ bidentate)[27,28] following the reaction below:

$$\mathrm{NO_{2(g)}} \leftrightarrow \mathrm{NO_{2(ads)}^{-}} + \mathrm{h^{+}}$$

Due to this charge transfer, the *c*MOF's electron density distribution is perturbed thus translating into detectable resistance changes. *c*MOF-based chemiresistors have thus been studied as a promising technology for $NO_2$ gas sensing.[24] To achieve sensor reversibility, approaches such as the use of elevated temperature, incorporation of noble metal catalysts, and photoactivation utilizing specific wavelength, and incorporation of heavy-metal nanoparticles have been reported to improve recovery kinetics, which still hamper real-world deployments.[21,22,24,29] In fact, these approaches remain the state-of-the-art, despite unique and promising features of *c*MOFs for $NO_2$ detection.[24,30] For real-world use, these approaches remain too costly, challenging to generalize, and user unfriendly, thus calling for innovative strategies to fully leverage the affinity of MOFs towards $NO_2$.

Efforts to expand the library of MOFs used in sensing applications have led to the utilization of polymer/MOFs hybrids when direct film growth of MOFs is challenging.[13–15] However, this approach often results in a compromise between processability and inherent properties. That is, MOFs are typically blended with polymer additives, which are often insulating polymers, thus masking the intrinsic properties of the MOF components.[13–15] Here we report a new concept to synergistically marry sensing performance and processability especially for detecting $NO_2$ gas at room temperature. We form hybrid films using designer conjugated polymers (*c*Ps) and conductive MOFs to improve: (i) The sensitivity compared to pristine *c*MOFs, stemming from increased density of the active material in the sensor. (ii) Chemical selectivity via the incorporation of systematically functionalized *c*Ps with selective and labile binding with $NO_2$. (iii) Regeneration of active sites within the sensing material, and consequently, long-term reliability at room temperature, due to improved kinetics of molecular exchange and a thermodynamically enhanced desorption process. And lastly, (iv) solution processibility via facile deployment methods (e.g., spin-coating, blade-coating, screen-printing, etc.), which is not readily unattainable in pristine MOFs.

## 2. Results and Discussion

### *2.1. Designing conductive polymer/MOF films for chemiresistive devices*

To form the newly designed polymer/MOF composite, we use a semicrystalline, mixed ionic-electronic conductive polymer (*c*P) based on 3,4-propylenedioxythiophene (ProDOT) and 2,1,3-benzothiadiazole (BTD),[31] and form hybrid films with 2D *c*MOFs (**Fig. 1 a-c** and **Supplementary Fig. 1, 2**). We select the ProDOT-BTD *c*P for its reversible redox-activity, low onset oxidation voltage, charge capacity, electrical conductivity to enhance the device response in presence of the analyte.[32,33] Particularly, the *c*P is designed to serve as a secondary $NO_2$-affine component endowed by its polar sidechains,[34–37] as well as a binding matrix to physically unify *c*MOFs crystallites and improve electrical communication throughout the bulk (**Fig. 1 c,d**). We then select two classes of *c*MOFs based on 2,3,6,7,10,11-hexahydroxytriphenylene (HHTP) and 2,3,6,7,10,11-hexaiminotripphenylene (HITP) ligands (**Fig. 1 a**), which typically exhibit irreversible gas sensing behaviors in their pristine form. As shown in **Fig. 1 e** and **Supplementary Fig. 2**, the x-ray diffraction peaks corresponding to pristine *c*MOFs remain unchanged upon hybridization without peak shifts, indicating that the structure of the *c*MOF remains undisturbed and well-preserved. We then study diverse metal nodes, which have shown promising performance in chemiresistive devices via molecular interactions between analyte gases and coordination sites.[38] All six *c*MOF combinations, namely $M_3(ligand)_2$, where M is either Co, Ni, or Cu and the ligand HHTP (X=O) or HITP (X=NH), were first synthesized and respective structures were verified through powder x-ray diffraction (PXRD) analysis (**Supplementary Fig. 2**).

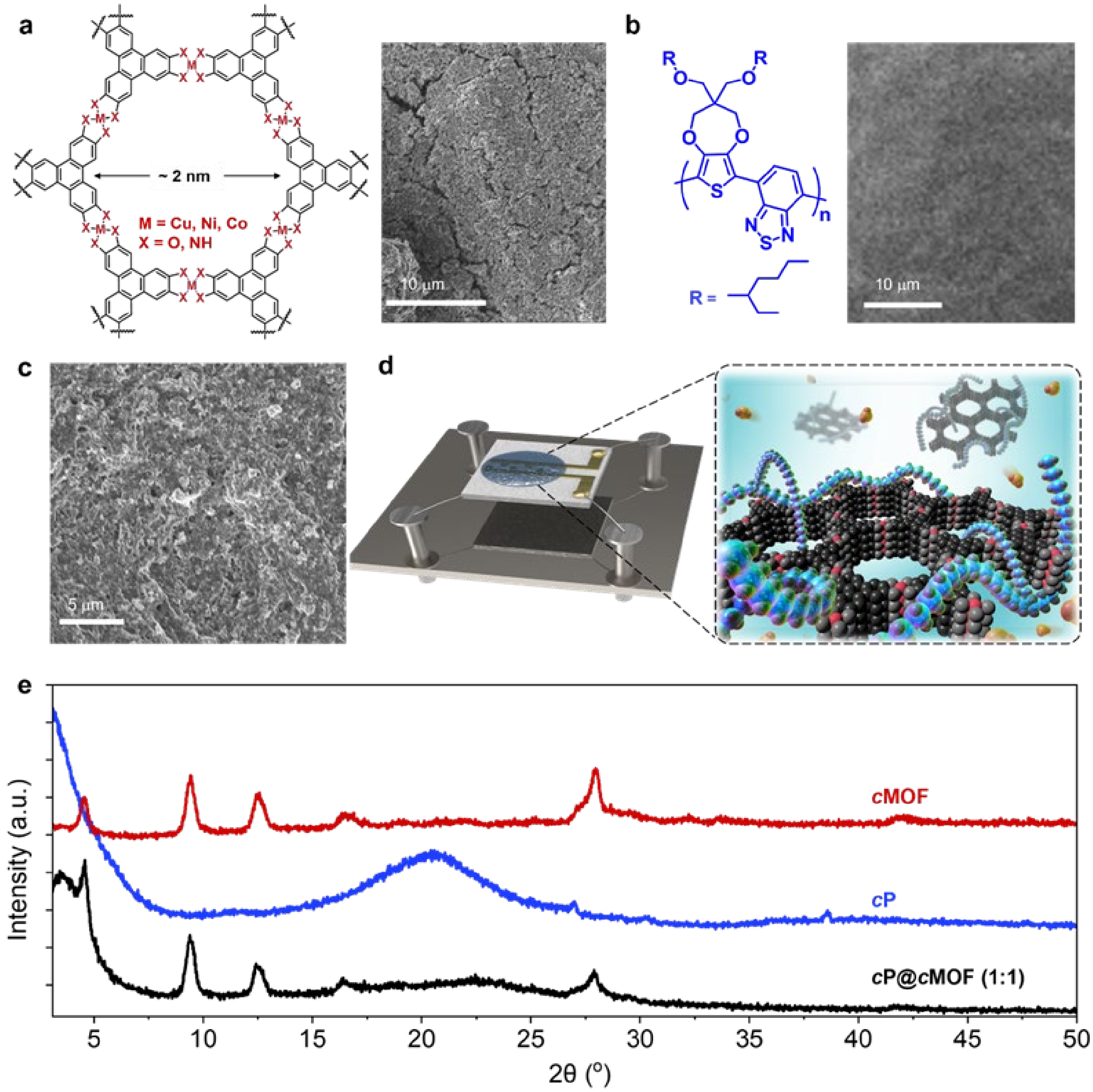

**Figure 1**. **Hybridization between conductive polymers and MOFs for robust chemiresistors.** (a), (b) Chemical structure and representative SEM images of 2D *c*MOFs, $M_3(ligand)_2$ (e.g., ($Cu_3HHTP_2$)) and *c*P. (c) SEM image of *c*P@*c*MOFs (1:1, w/w). (d) Schematic illustration of sensing device and interaction between *c*P@*c*MOF and gas analyte. (e) PXRD spectra of 2D *c*MOF ($Cu_3HHTP_2$), *c*P, and corresponding *c*P@*c*MOFs (1:1, w/w).

Scanning electron microscope (SEM) images show that the hybrid *c*P@*c*MOFs films are homogeneous and *c*MOFs crystallites are uniformly embedded in the *c*P matrix (e.g., *c*P@$Cu_3HHTP_2$, **Fig. 1 c**). This is in contrast with pristine *c*MOF films which show micro-cracks and pristine *c*P with a smooth texture (**Fig. 1 a,b**). We also use atomic force microscopy (AFM) imaging to compare surface roughness of drop-casted films, revealing smoother surfaces in

*c*P@*c*MOFs compared to pristine *c*MOFs with Rq values of 97.5 nm and 195 nm, respectively, owing to the presence of the *c*P with Rq value of 22.7 nm (**Supplementary Fig. 3a-c**). Furthermore, as corroborated by surface analysis using x-ray photoelectron spectroscopy (XPS) and Raman spectroscopy (probing depths of 10 nm and 645 nm, respectively), the *c*P shows to constitute most of the outermost layer of the composite films (at least ~10 nm), engulfing *c*MOFs crystallites. At such shallow depths, the distinctive metal peaks from the *c*MOF component were absent in high-resolution XPS spectra of hybrid films (**Supplementary Fig. 4**). At the same time, corresponding Raman spectra were identical to that of pristine *c*P (**Supplementary Fig. 5**) indicative of a predominantly polymer-enriched interface in the hybrid films. We hypothesize that with this architecture, the semicrystalline and $NO_2$-affine polymer would also contribute to the adsorption of gas molecules into sensor's channel area and enhance the overall sensitivity of the sensors. That is, i) the conductive backbone is flanked with polar sidechains favoring the permeability of gas molecules into the active area across entire active channel,[29-32] and ii) *c*P serves as a conductive matrix binding together the MOF crystallites throughout the bulk, thus promoting efficient charge transport upon resistance change. By interlinking the crystallites throughout the film bulk, the *c*P thus helps reduce inter-particles resistance, a dominant behavior in pristine MOFs, and enhances the performance of chemiresistors based on polymer/MOF composites (**Fig. 1 c,d**). The pasty nature of the solution processed *c*P was also envisioned to promote greater films integrity and hence device reliability and stability.

### *2.2. Gas sensing performance of cP@cMOFs*

To test the effect of hybridization on gas sensing performance, we fabricated chemiresistive sensors using the *c*P@*c*MOFs combinations discussed above. Further details on the sensor fabrication steps and device dimensions are provided in the experimental section and illustrated in **Supplementary Fig. 6**. As shown in **Fig. 2 a**, all sensors based on pristine *c*MOFs exhibited relatively low responses ($R_{air}/R_{gas}$ or $R_{gas}/R_{air}$ < ~2, where $R_{air}$ and $R_{gas}$ denote resistance in air and gas, respectively) as well as poor sensing reversibility. Upon hybridization with *c*P (e.g., 1:1 weight ratio between *c*P and *c*MOF), the sensing response, and most importantly, the sensing reversibility was significantly enhanced across all *c*MOFs (**Fig. 2 a-c** and **Supplementary Fig. 7**). This performance enhancement was most exemplified in *c*P@$Ni_3(HITP)_2$ exhibiting a 23.9-fold improvement in sensor response relative to pristine $Ni_3(HITP)_2$ (**Fig. 2 c**). Note that, in its pristine form the *c*P yields sensors with undetectable response, and the same effect could not be induced when the hybridization is done with corresponding ligands (HHTP or HITP) instead of *c*MOFs, or other commonly studied conductive materials such as carbon nanotubes (CNTs) (**Supplementary Figs. 8, 9)**. This behavior thus underscores the importance of utilizing a conductive polymer to hybridize with the porous *c*MOFs and synergistically enhance sensing response and reversibility.

Furthermore, the hybridization was beneficial to the sensitivity level of the sensor devices as illustrated in the dynamic resistance changes with $NO_2$ concentrations down to 0.25 ppm (**Fig. 2 d-f**). In terms of sensitivity, we found that, upon optimization of the *c*P content in the hybrid films (**Supplementary Figs. 10-15)**, *c*P@$Ni_3(HITP)_2$ demonstrated the most pronounced initial response in the presence of $NO_2$, albeit its relatively modest reversibility compared to the *c*P@$Co_3(HHTP)_2$ analogue. Also noteworthy, among all hybrid combinations, *c*P@$Co_3(HHTP)_2$ exhibited the most dynamic resistive behavior and reversibility enhancement in comparison to its pristine counterpart. Nonetheless, across all six *c*MOFs, the hybridization showed to enhance the sensing response compared to pristine constituents (**Fig. 2 g**) while conserving the selectivity level towards $NO_2$ gas (**Supplementary Figs. 16, 17**), given uniform distribution of the components within the bulk (**Supplementary Fig. 18**). Most importantly, all *c*P@*c*MOFs-based sensors demonstrated enhanced cycling reversibility compared to their pristine counterparts (**Fig. 2 h-i** and **Supplementary Figs. 10,14**). Particularly, *c*P@$Co_3(HHTP)_2$-based sensors exhibit stable and reversible performance up to 97 cycles (**Fig. 2 h-i**). To the best of our knowledge, our work presents the highest number of cycling tests with stable reversibility among all *c*MOFs- or *c*P-based chemiresistive sensors reported to date (**Table 1**).[21]

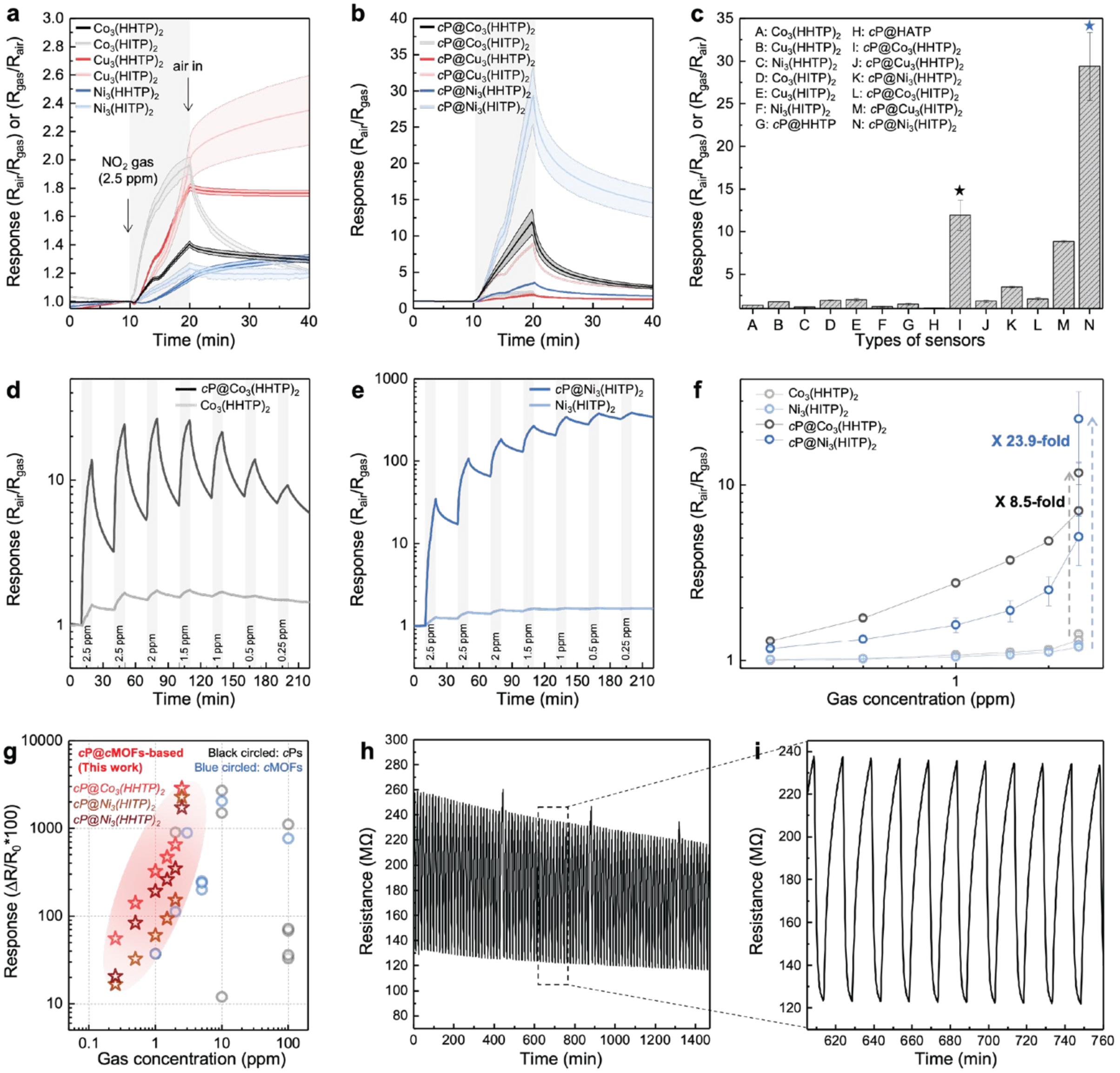

**Figure 2**. **Sensing performance evaluation in *c*P@*c*MOF-based chemiresistors.** Response graphs of (a) 6 different pristine *c*MOFs and (b) *c*P@*c*MOFs with 1:1 ratio. (c) Response of all the sensors including pristine *c*MOFs, *c*P@ligands, and *c*P@*c*MOFs (N ≥ 3). (d) Response graphs of $Co_3(HHTP)_2$ and *c*P@$Co_3(HHTP)_2$ toward 2.5-0.25 ppm $NO_2$ gas. (e) Response graphs of $Ni_3(HITP)_2$ and *c*P@$Ni_3(HITP)_2$ toward 2.5-0.25 ppm $NO_2$ gas. (f) Response of $Co_3(HHTP)_2$, $Ni_3(HITP)_2$, *c*P@$Co_3(HHTP)_2$, and *c*P@$Ni_3(HITP)_2$ toward 2.5-0.25 ppm $NO_2$ gas (N ≥ 3). (g) Responses of reported $NO_2$ sensors using *c*P or *c*MOFs, operating at RT.[19,21,22,29,39–50] (h) Operational stability of a *c*P@$Co_3(HHTP)_2$ sensor under 97 cyclic exposures (1 ppm $NO_2$, 5 min exposure, 10 min air recovery), including (i) an enlargement around t = 605-760 min.

| Sensing material | Response [ΔR/R$_0$ * 100 %] | Cyclability (# of cycles) | LOD [ppm] | Ref. |
|---|---|---|---|---|
| PPy thin film | 12 @ 10 ppm | 3 | 10 | 39 |
| PT thin film | 33 @ 100 ppm | 3 | 10 | 40 |
| PANI-nanofibers | 1500 @ 10 ppm | n.r. | 10 | 41 |
| PANI thin film | 1110 @ 100 ppm | n.r. | 10 | 42 |
| Ag-PPy | 68 @ 100 ppm | n.r. | 5 | 43 |
| PANI fibers | 900 @ 2 ppm | n.r. | 1 | 44 |
| ZnO/PANI nanoflake | 2700 @ 10 ppm | n.r. | 0.01 | 45 |
| DBSA-doped PPy–$WO_3$ | 72 @ 100 ppm | 3 | 5 | 46 |
| PPy/α-$Fe_2O_3$ | 36 @ 100 ppm | 3 | 10 | 47 |
| PEI-doped PNDIT2/IM | 37.2 @ 1 ppm | 15 | 0.1 | 48 |
| $Cu_3(HHTP)_2/Fe_2O_3$ hybrids | ~200 @ 5 ppm | 7 | 0.2 | 21 |
| Pd@$Cu_3(HHTP)_2$ | ~250 @ 5 ppm | 14 | 1 | 22 |
| Pt@$Cu_3(HHTP)_2$ | ~240 @ 5 ppm | 14 | 1 | 19 |
| PtRu@$Cu_3(HHTP)_2$ | 112.8 @ 2 ppm | 7 | 0.2 | 19 |
| Pt@$Cu_3(HHTP)_2$ thin film | 890.1 @ 3 ppm | n.r. | 0.1 | 26 |
| Cu-Salphen-MOF | 766 @ 100 ppm | 5 | 1 | 50 |
| Single crystal Ti-MOF (FIR-120) | 2040 @ 10 ppm | 5 | 1 | 29 |
| ***c*P@CoHHTP 1:4** | **2863 @ 2.5 ppm** | **97** | **0.25** | **This work** |
| ***c*P@NiHITP 1:1** | **2282 @ 2.5 ppm** | **10** | **0.25** | **This work** |
| ***c*P@NiHHTP 8:1** | **1716 @ 2.5 ppm** | **10** | **0.25** | **This work** |

**Table 1.** Comparison of current results to state-of-the-art $NO_2$ sensing at room temperature using *c*Ps- and *c*MOFs-based chemiresistors (n.r.: not reported).

### *2.3. The role of hole enrichment toward reversible $NO_2$ sensing*

Key to synergistically deploying both conductors in our hybrid films is the electronic characteristic at the heterojunctions formed between *c*P and *c*MOFs, as well as the distribution of such heterojunctions throughout the bulk. Particularly, the activation energy for charge carrier transport in response to $NO_2$ adsorption is pivotal for enhancing the sensing performance of hybrid films. Here we sought to evaluate the sensor reversibility at room temperature by monitoring the channel resistance recovery before and after $NO_2$ exposure. As discussed above and confirmed by solid state characterizations, the structural configuration of the hybrid film (i.e., lowered film crystallinity and uniform distribution of *c*MOF crystallites within the *c*P bulk) accounts for enhancement in sensitivity and its retention. Given that our *c*P@*c*MOFs-based sensors also exhibit excellent signal recovery, we sought to mechanistically understand the impact of these features on the adsorption and desorption kinetics of $NO_2$ gas. In MOF-based $NO_2$ sensors, reducing the sensor response times and, more importantly, increasing the desorption rate constant ($k_{des}$) to achieve reversible sensing has been challenging.[26]

Without relying on external stimuli or addition of inactive components, the newly designed *c*P@*c*MOFs hybridization provides a thermodynamic solution to irreversible detection of $NO_2$. The experimentally constructed energy band diagrams showing respective HOMO levels with respect to fermi level reveal p-type *c*P as the most hole-rich component (**Fig. 3 a** and **Supplementary Figs. 19, 20** and **Table S1**). In all studied *c*MOF cases, the two materials are energetically close enough allowing us to hypothesize that, upon hybridization, beyond the formation of microscopic interfaces throughout the bulk, the *c*P-*c*MOF heterojunctions establish an electronic equilibrium, and the majority carriers occupy a shared Fermi level (**Fig. 3 a**). In other words, a hole transfer from the *c*P to the *c*MOF is thermodynamically favorable forming a hybrid and hole-enriched *c*MOF state (**Fig. 3 a** and **Supplementary Figs. 20**, **21**).[51,52] This injection of holes into the *c*MOF's electronic configuration alters the interaction between $NO_2$ and the sensing channel by lowering the binding energy (the primary source of irreversible sensing, **Supplementary Fig. 22**), resulting in reversible sensing behavior, even at room temperature (**Fig. 3 a**). Prior to this work, the desorption of $NO_2$ gas from MOF sorbate materials has been achieved using elevated temperatures, high energy radiation, or the incorporation of heavy metals in the active layer.[26] The hybridization strategy we report here is much more straightforward and holds potential for generalization onto essentially any conductive MOF structure.

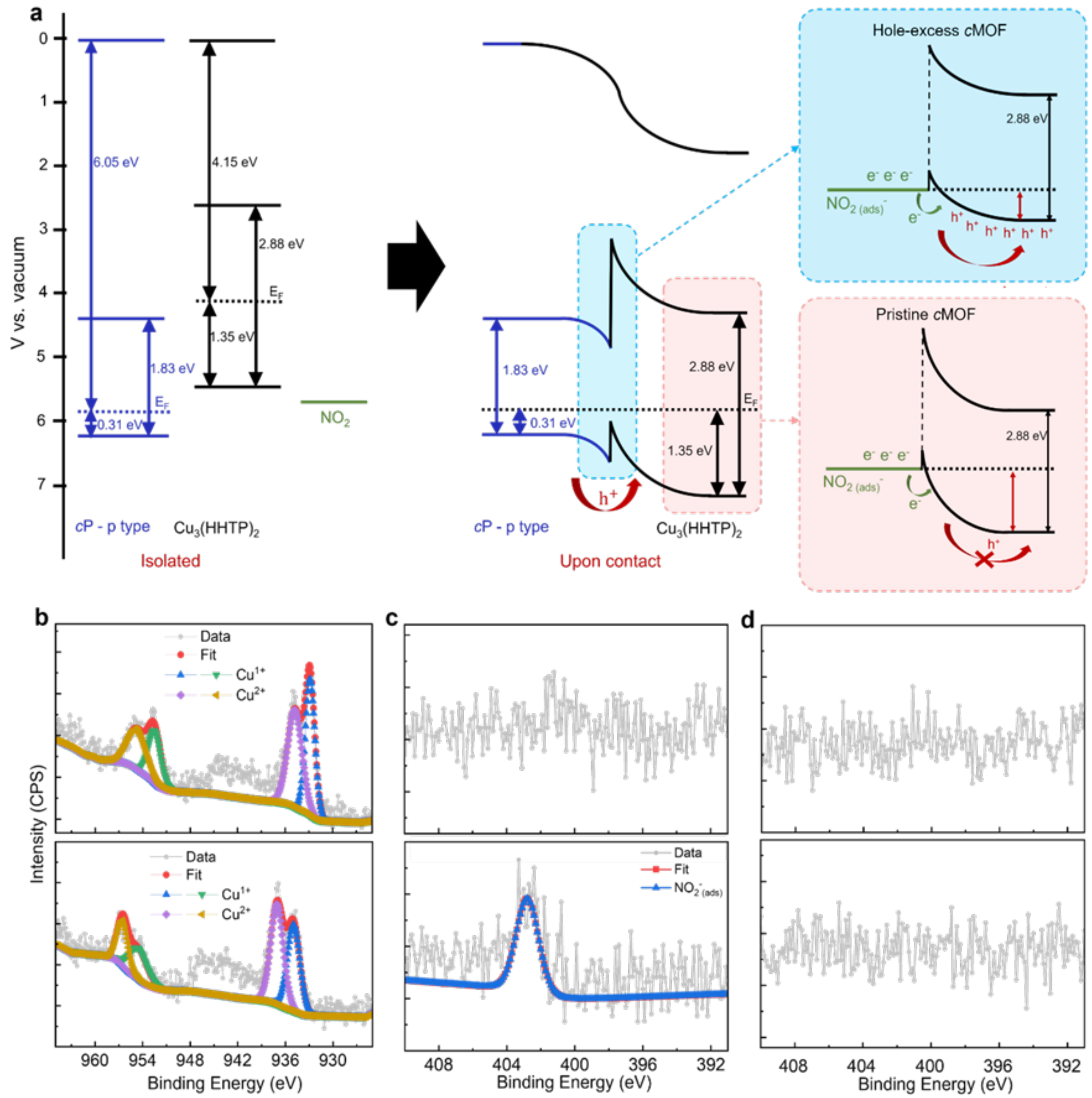

**Figure 3. Hole enrichment in *c*MOFs and its impact on analyte binding.** (a) Experimentally constructed energy level diagram of *c*P and representative *c*MOF (full data for all *c*MOFs studied can be found in **Supplementary Figs. 19,20**) and proposed mechanism/rationale for enhanced recovery upon hybrid. (b) High resolution XPS spectra of Cu 2p peak region, before (top) and after (bottom) $NO_2$ exposure. (c, d) High resolution XPS spectra of N 1s peak region of (c) $Cu_3(HHTP)_2$ and (d) *c*P@$Cu_3(HHTP)_2$, respectively, before (top) and after (bottom) $NO_2$ exposure.

To test our proposed working mechanism, we first examined the XPS spectra of the active material before and after $NO_2$ exposure. The oxidative effect of $NO_2$ on the sensing materials was evidenced by subsequent charge redistribution around the metal nodes to compensate the charge imbalance. High-resolution XPS spectra of the characteristic metal peaks (e.g., Cu 2p) reveal a significant change in the ratio between oxidation states of the metal (e.g., $Cu^{1+}$ and $Cu^{2+}$). In the case of $Cu_3(HHTP)_2$ films, substantial change in the oxidation state of the Cu was evidenced by a decrease in the $Cu^{1+}$ (~933 eV)/$Cu^{2+}$ (~935 eV) ratio from 1.17 to 1.09 after exposure to $NO_2$ (**Fig. 3 b**). Concomitantly, in the pristine MOF samples, despite the high vacuum conditions during XPS measurements, the adsorption and binding of $NO_2$ molecules was evidenced by a signature peak at 403.9 eV (**Fig. 3 c**). In contrast, the hybrid films exhibit no discernable signal from this N 1s peak within the same desorption timescale (**Fig. 3 d**), indicative of nearly complete desorption of $NO_2$ gas from the hybrid films. We associate this complete desorption to the formation of a new Fermi level upon hole enrichment, thus weakened binding between the gas molecules and the *c*MOF sites. This labile binding between $NO_2$ molecules and our *c*P@*c*MOFs films would thus be the key rationale to the observed dynamic response in the chemiresistive sensors.

We further corroborated this sensing mechanism by experimentally monitoring the sensor's recovery kinetics upon hybridization. We employed a mass action law of gas adsorption reactions on both *c*MOFs and *c*P@*c*MOFs and computed the response and recovery kinetics for $NO_2$ sensing. Our calculations assumed that the quantity of gas adsorbed on the surface is directly related to the sensors' response. We then calculated respective rate constants according to previous works (i.e., $k_{ads}$ for gas adsorption and $k_{des}$ for the desorption)[53,54] by fitting the sensor's response graphs using equations (1) and (2) below for six distinct *c*MOFs and corresponding *c*P@*c*MOFs (**Supplementary Fig. 23**):

$$\text{R(t) for NO}_2\text{ adsorption} = \mathrm{R_{max}} \cdot \frac{\mathrm{C_a K}}{1+\mathrm{C_a K}}\left(1 - \exp\left[-\frac{1+\mathrm{C_g K}}{\mathrm{K}} \cdot \mathrm{k_{ads}} t\right]\right), \quad (1)$$

$$\text{R(t) for NO}_2\text{ desorption} = \mathrm{R_0} \exp[-\mathrm{k_{des}} t]. \quad (2)$$

Here, $R_0$ is the baseline response in air, $R_{max}$ is the maximum response, $C_a$ is the gas concentration, t is time, and K is an equilibrium constant ($k_{ads}/k_{des}$).[53] In all cases, *c*P@*c*MOFs displayed remarkably improved desorption kinetics compared to their corresponding pristine *c*MOFs counterparts. Notably, the $k_{des}$ for *c*P@$Cu_3(HITP)_2$ was measured to be 93.7-fold higher compared to that of pristine $Cu_3(HITP)_2$. Interestingly, the $k_{ads}$ values for all *c*P@*c*MOFs exhibited minimal changes when compared to those of pristine *c*MOFs. It was thus evident that the dominant factor for enhancing reversibility in the composite systems is the thermodynamic effect from the hole enrichment, rather than structural factors. Note that for $Cu_3(HITP)_2$ and $Ni_3(HHTP)_2$, which showed n-type resistive variation upon $NO_2$ exposure (oxidizing gas), a higher amount of *c*P was required than other *c*MOFs to achieve the optimal ratio for reversibility (**Supplementary Figs. 10-12, 14**). Serendipitously, these two systems also exhibit significantly larger ($E_F$ - $E_V$) values relative to the *c*P in energy level diagram, making the hole enrichment more energy consuming. With the same rationale, superior room temperature reversibility was observed in pristine Co-based *c*MOFs compared to other pristine *c*MOFs and could be attributed to their notably lower ($E_F$ - $E_V$) value (~0.5 eV) (thus inherent abundance of hole carrier density), set aside lower crystallinity (**Supplementary Fig. 24**).

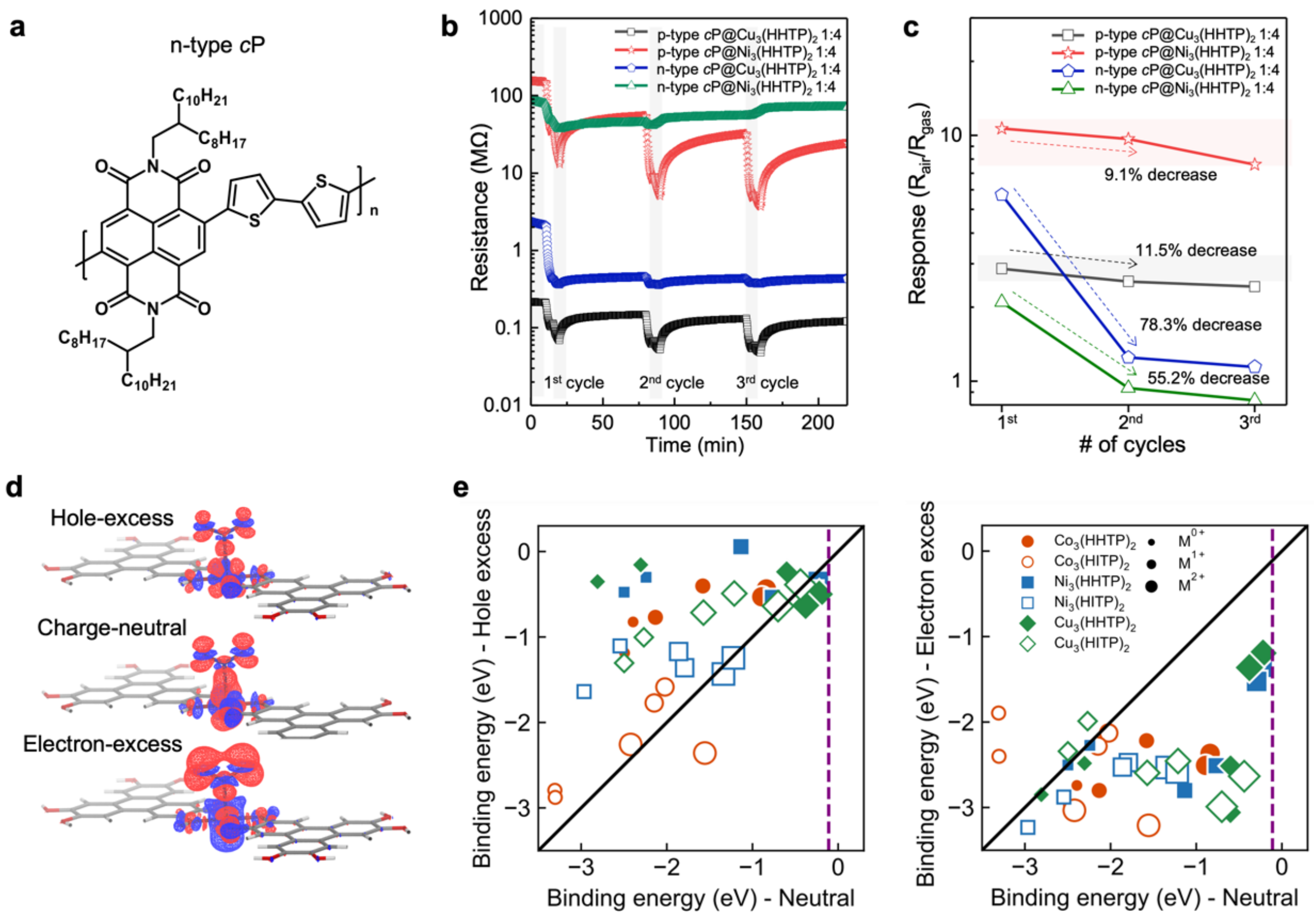

**Figure 4. Relationship between binding energy and sensor performance.** (a) Chemical structure of n-type polymer studied. (b) Dynamic resistance transitions with varied types of polymers in hybrid. (c) Response profile with varied types of polymers in hybrid. (d) Differences in charge density before and after $NO_2$ binding to the *c*MOF cluster for three system conditions: hole-excess, charge-neutral, and electron-excess, calculated as ρ (Cluster + $NO_2$) - ρ (Cluster) - ρ ($NO_2$), where ρ is the charge density. The red surface denotes an increase in charge density upon $NO_2$ binding, while the blue surface denotes a decrease. An iso-surface value of 0.0025 $e^{-1}/Å^3$ has been applied. (e) Binding energy of *c*MOF and $NO_2$ when the *c*MOF has an excess hole (left)

and an excess electron (right), compared to charge-neutral *c*MOF. Both adsorption cases, where $NO_2$ binds either through the O or N atom, are illustrated. The purple dashed line represents the average binding energy between the *c*P and $NO_2$. The black solid line indicates parity.

To verify the role of hole enrichment and test potential countereffect from electron enrichment, we substituted the *c*P component in our hybrid films with an n-type polymer (in this case, poly{[N,N′-bis(2-octyldodecyl)-naphthalene-1,4,5,8-bis(dicarboximide)-2,6-diyl]-alt-5,5′-(2,2′-bithiophene)}, also known as N2200) (**Fig. 4 a**). We then fabricated four types of sensors, i.e., p-type *c*P@$Cu_3$(HHTP)$_2$, p-type *c*P@$Ni_3$(HHTP)$_2$, n-type *c*P@$Cu_3$(HHTP)$_2$, and n-type *c*P@$Ni_3$(HHTP)$_2$, and compared their sensing behavior under similar conditions. More specifically, we monitored the channel resistance recovery after 3 cyclic exposures of the sensor to 2.5 ppm $NO_2$. P-type *c*P-based composites exhibited reversible sensing behaviors with minimal reduction in response, 9.1% and 11.5%, respectively, after the first cycle $NO_2$ exposure (**Fig. 4 b,c**). Conversely, n-type *c*P-based composites demonstrated a dramatically diminished response after the first cycle. Furthermore, exposure to higher $NO_2$ concentrations in subsequent cycles showed no further change in the channel resistance for n-type composites (**Supplementary Fig. 25**) indicative of a dosimetric sensing behavior. For instance, n-type *c*P@$Cu_3$(HHTP)$_2$ exhibited a higher response during the initial cycle, attributed to its abundant electron concentration facilitating electron donation to $NO_2$, but this enhanced response rapidly diminished by 78.3% during the second cycle, attributable to the incomplete recovery (or irreversible kinetics) (**Fig. 4 c**). These results effectively emphasize the significance of hole enrichment in the hybrid films enabled by the polymeric component and its effect on the desorption of $NO_2$ gas, promoting favorable recovery kinetics.

Using density functional theory (DFT) calculations, we also calculated the binding energies between the gas molecules and the sensing material to gain further insight on the effect of *c*P hybridization on the observed sensing behaviors.[55,56] For our simulations, instead of modeling the entire *c*P@*c*MOFs systems, we represented *c*P@*c*MOFs with p/n-type *c*P as pure *c*MOFs with an additional injected hole/electron. By comparing the hybrid systems with the pure, charge-neutral *c*MOFs, this approach allows us to focus on the hole/electron enrichment effect of *c*Ps on *c*MOFs. For all six *c*MOFs under consideration, our simulations considered three possible metal oxidation states, +0, +1, and +2 on the metal node (**Fig. 4 d** and **Supplementary Fig. 26**). Across all *c*MOFs, we observed that hole-enriched states exhibit lower binding energy than their neutral state and electron-rich states (**Fig. 4 e**). These simulation results thus further supported our rationale of hole-enrichment as a major contributor to $NO_2$ desorption and thus sensing reversibility. Note that the case of Co-based *c*MOFs discussed above also show binding energies on par with other *c*MOFs, thus underscoring the importance of other contributing factors to the sensing kinetics such as film crystallinity and crystallite size (**Supplementary Fig. 24 c,d**). We could thus conclude that, in hole-excess scenarios, less electronic charge is transferred from the metal nodes to $NO_2$, thereby weakening their binding, compared to electron-excess systems (**Fig. 4 b**). This charge transfer between the $NO_2$ and the metal nodes also supports superior selectivity over other gases, such as $H_2S$ and $NH_3$, as evidenced by calculated selectivity data in **Supplementary Figs. 22, 27**.

## 3. Conclusions

In conclusion, we propose a hybridization method combining two complementary classes of mixed ionic-electronic conductors (MIECs), conductive polymers and MOFs (*c*P and *c*MOFs). Our approach produces chemiresistors with better cycling stability and sustained dynamic range than those made solely from *c*MOFs or *c*Ps. We conducted a detailed analysis to understand how hybridization with the conductive polymers enhances the reversibility of *c*MOFs-based sensors, focusing on energy band alignments at the material interfaces and their impact on sensing thermodynamics and binding kinetics. Theoretical calculations further elucidated the effect of such hybridization of the interaction between *c*MOFs and the gaseous analyte. Our approach proves to be versatile towards designing conductive polymer/MOF composites with improved performance and processability. The hybridization thus paves way for more tailored composite-based electronics leveraging the intrinsic properties of both polymers and MOFs.

## 4. Materials and Methods

### *Materials*

$Co(OAc)_2 \cdot 4H_2O$ (Alfa Aesar), $Co(NO_3)_2 \cdot 6H_2O$ (Alfa Aesar), $Cu(OAc)_2 \cdot xH_2O$ (Alfa Aesar), $CuSO_4 \cdot 5H_2O$ (Alfa Aesar), $Ni(OAc)_2 \cdot 4H_2O$ (Strem), and NaOAc (Alfa Aesar) were used without further purification. 2,3,6,7,10,11-hexaaminotriphenylene hexahydrochloride (HATP·6HCl) was prepared according to a procedure published elsewhere.[57] 2,3,6,7,10,11-hexahydroxytriphenylene hydrate (HHTP, $C_{18}H_{12}O_6 \cdot H_2O$, 95%) was purchased from Tokyo Chemical Industry. For the synthesis of HHTP with higher crystallinity, recrystallization of HHTP ligand was conducted. N,N-dimethylformamide (DMF), acetone, and methanol were used as received without further purification. Further synthetic details can be found in Supplementary Information.

***cP@cMOF hybrid film processing***

A polymer solution of 10 mg/ml in chloroform was prepared and stirred for 30 minutes at 35 °C. The *c*MOFs were initially dispersed in deionized water and ethanol, for $M_3(HITP)_2$ series and $M_3(HHTP)_2$ series, respectively, immediately after synthesis and filtration. Subsequently, the dispersed *c*MOFs were sonicated for 5-30 minutes and blended with the polymer solution in appropriate w/w ratio to create the final solution for the hybrid film.

***Powder X-ray diffraction (PXRD)***

PXRD analysis was conducted using a Bruker Advance II diffractometer equipped with a θ/2θ reflection geometry and Ni-filtered Cu Kα radiation ($K_{\alpha1}$ = 1.5406 Å, $K_{\alpha2}$ = 1.5444 Å, $K_{\alpha2}/K_{\alpha1}$ = 0.5). The tube voltage and current were set as 40 kV and 40 mA, respectively, during operation. Samples were prepared by placing the material on a zero-background silicon crystal plate.

***Spectroscopy measurements***

X-ray photoelectron spectroscopy (XPS) measurements were performed using a Physical Electronics PHI Versaprobe II X-ray photoelectron spectrometer equipped with an Al anode as a source. To prepare the samples for analysis, powders were compressed onto copper tapes to ensure complete coverage. The calibration of charge shift was performed by aligning the C1s peak of surface-adsorbed adventitious carbon to 284.8 eV. Ultraviolet-visible (UV-vis) absorption spectra were acquired using a Perkin Elmer 1050 UV-visible-NIR spectrophotometer. Raman spectra were collected on a Raman Reflex instrument utilizing a 532 nm laser source.

***Chemiresistor fabrication and characterization***

Gas sensing characterizations involved the coating of sensing materials onto prepatterned $Al_2O_3$-based sensor substrates. These substrates featured two parallel electrodes, each measuring 90 µm in width and spaced 160 μm apart. Following the preparation of the sensing material solutions at a concentration of 10 mg/ml, 5 μl of the solution were cast onto the sensor substrate and dried. Measurement of the resistance of the sensing materials on the electrodes was carried out using a data acquisition system (Agilent 34972A) equipped with a 16-channel multiplexer (Agilent 34902A). For evaluation of sensing characteristics, the measured resistance values were converted into response values ($R_{air}/R_{gas}$ or $R_{gas}/R_{air}$), where $R_{air}$ and $R_{gas}$ denoted the resistance in air and the gas, respectively. For stabilization of the baseline resistance, a baseline air was employed to stabilize the sensors for at least 2 hours. To establish concentration-dependent measurements, gas cylinders containing 50 ppm $NO_2$, 50 ppm $H_2S$, were purchased from Airgas. These gases were then diluted with air using mass flow controllers. Other organic analytes including ethanol, methanol, acetone, toluene, and xylene, were supplied to the sensing chamber using a FlexStream FlexBase module. Further details regarding sensor fabrication steps and sensor dimensions are provided in Supplementary Information.

***Experimental energy level measurement***

The electronic band structures were constructed collectively by following measurements: X-ray photoelectron spectroscopy (XPS) with monochromatic Al Ka = 1486.6 eV with -10 V bias was conducted to obtain both cut-off and fermi spectra. A gentle ion gun treatment (Monatomic source gun with 2000 eV, 30 seconds etch time) was performed to clean the surface. Tauc plots were converted from UV-vis spectra.

***Theoretical calculation details: density functional theory (DFT) calculations***

To enable facile control of the charge and oxidation state in our DFT studies, we studied a finite cluster model of the *c*MOF system using the hybrid B3LYP functional and the composite LACVP* basis set. These cluster models were extracted from DFT-optimized structures of the full periodic *c*MOFs with PBE-D2 and a plane wave basis set (kinetic energy cutoff of 400 eV). Full computational details are provided in the Supplementary Information.

**Acknowledgements**

H.R. and D.-H.K. contributed equally to this work. H. R. and A. G. acknowledge financial support from the MIT Climate & Sustainability Consortium (MCSC) as well as the Abdul Latif Jameel

Water & Food Systems (JWAFS) Lab at the Massachusetts Institute of Technology (MIT). D. H-K. and M. D. acknowledge support from the U.S. Department Of Energy (DE-SC0023288). Y. C. and H. J. K. acknowledge support for the U.S. Department Of Energy (DE-SC0023288) DE-NA0003965.

Supporting Information

**Robust Chemiresistive Behavior in Conductive Polymer/MOF Composites**

*Heejung Roh, Dong-Ha Kim, Yeongsu Cho, Young-Moo Jo, Jesús A. del Alamo, Heather J. Kulik,*Mircea Dincă,* Aristide Gumyusenge **

**Table of Contents**

## 1) Synthetic Details

***Synthesis of $Co_3(HHTP)_2$***

20 mg of $Co(OAc)_2 \cdot 4H_2O$ was dispersed in 4 ml of water. Then, 16.2 mg of recrystallized HHTP ligand was dispersed in the mixture of 4 ml of water and 2 ml of DMF solvents. After mixing of the two solutions, sonication was conducted for 5 minutes. Then, the mixed solution was put into the sand heated at 85 °C for 24 hours (vial closed, without stirring). After reaction, the solution was filtered and washed with a large amount of water and acetone. Then, the obtained power was dried overnight. Note that for the synthesis of $Co_3(HHTP)_2$ with higher crystallinity, 16.2 mg of recrystallized HHTP ligand was dispersed in the mixture of 1.33 ml of water and 0.67 ml of DMF solvents. All the other synthetic procedures remain unchanged.

***Synthesis of $Cu_3(HHTP)_2$***

24.9 mg of $Cu(OAc)_2 \cdot H_2O$ was dispersed in 4 ml of water. 16.2 mg of recrystallized HHTP ligand was dispersed in 4 ml of DMF solvent. After mixing of the solutions, sonication was conducted for 5 minutes. Then, the mixed solution was put into the sand heated at 85 °C for 24 hours (vial closed, without stirring). After reaction, the solution was filtered and washed with a large amount of water and acetone. Then, the obtained power was dried overnight. For the synthesis of $Cu_3(HHTP)_2$ with lower crystallinity, commercial HHTP ligand (without recrystallization process) was utilized with the same synthetic protocols.

***Synthesis of $Ni_3(HHTP)_2$***

24.9 mg of $Ni(OAc)_2 \cdot 4H_2O$ was dispersed in 4 ml of water. 16.2 mg of recrystallized HHTP ligand was dispersed in 4 ml of water. After mixing the solutions, sonication was conducted for 5 minutes. Then, the mixed solution was put into the sand heated at 85 °C for 24 hours (vial closed, without stirring). After reaction, the solution was filtered and washed with a large amount of water and acetone. Then, the obtained power was dried overnight.

***Synthesis of $Co_3(HITP)_2$***

8.13 mg of $Co(NO_3)_2 \cdot 6H_2O$ was dispersed in 3 ml of DMF. 10 mg of $HATP \cdot 6HCl$ ligand was dispersed in 3 ml of water. After mixing the solutions, sonication was conducted for 5 minutes. Then, 4 ml of 2M NaOAc was added to the solution and the mixed solution was put into the sand heated at 65 °C for 2 hours (vial opened, with stirring). After reaction, the solution was filtered and washed with a large amount of water and methanol. Then, the obtained power was dried overnight.

***Synthesis of $Cu_3(HITP)_2$***

7 mg of $Cu(SO_4)_2{\cdot}5H_2O$ was dispersed in 3 ml of DMF. 10 mg of HATP·6HCl ligand was dispersed in 3 ml of water. The mixed solution was sonicated for 5 minutes and put into the sand heated at 65 °C. Then, 4 ml of 2M NaOAc was added to the solution and heated at 65 °C for 2 hours (vial opened, with stirring). After reaction, the solution was filtered and washed with a large amount of water and methanol. Then, the obtained power was dried overnight.

***Synthesis of $Ni_3(HITP)_2$***

6.94 mg of $Ni(OAc)_2{\cdot}4H_2O$ was dispersed in 3 ml of DMF. 10 mg of HATP·6HCl ligand was dispersed in 3 ml of water. The mixed solution was sonicated for 5 minutes and put into the sand heated at 65 °C. Then, 8 ml of 2M NaOAc was added to the solution and heated at 65 °C for 2 hours (vial opened, with stirring). After reaction, the solution was filtered and washed with a large amount of water and methanol. Then, the obtained power was dried overnight.

***Synthesis of P-type cP***

P-type *c*P selected for this study consists of ProDOT and BTD building units according to the previous reported procedure[1].

***Synthesis of N-type cP***

N-type *c*P selected for this study is N2200, or PNDI-2T, or Poly{[N,N′-bis(2-octyldodecyl)-naphthalene-1,4,5,8-bis(dicarboximide)-2,6-diyl]-alt-5,5′-(2,2′-bithiophene)[2] was synthesized using a Pd-catalyzed Stille coupling reaction. $NDI\text{-}Br_2$ (0.50 g, 0.46 mmol) and 5,5'-bis(trimethylstannyl)-2,2'-bithiophene (0.226 g, 0.46 mmol) were dissolved in dry chlorobenzene (7.5 mL). After degassing with $N_2$ for 1 h, $Pd_2(dba)_3$ (8 mg) and $P(o\text{-}Tol)_3$ (11 mg) were added to the mixture and stirred for 48 h at 110 °C. Subsequently, 2-bromothiophene and tributyl(thiophen-2-yl)stannane were injected to the reaction mixture for end-capping, and the reaction was stirred for 6 h. The polymer was precipitated in methanol, collected by filtration, and then purified by successive Soxhlet extraction with methanol, acetone, hexane, toluene, and chloroform. The final product was obtained by precipitation in methanol.

## 2) Density Functional Theory (DFT) Calculations Details

Our starting structure was based on the experimentally determined lattice parameters of $Cu_3(HHTP)_2$ as reported in reference 1[3]. Each unit cell includes two layers, which were offset by a fractional coordinate of (1/32, 0, 0), following reference 2[4] **(Supplementary Fig. 25a,b)**. The atomic positions and lattice parameters were optimized using the Vienna Ab Initio Simulation Package (VASP) version 6.3.1[5-8], until the total energy converged to within 0.01 eV. We employed the Perdew-Burke-Ernzerhof (PBE) functional[7] along with the plane-augmented wave (PAW) pseudopotential[10]. A kinetic energy cutoff of 400 eV was used, and DFT-D2 corrections were applied to account for dispersion interactions[9]. Given the large size of the unit cell, calculations were performed at the Γ point only. A background charge of +12 per unit cell was applied to the system. Additionally, each metal atom was assumed to have an oxidation state of +2 and to be in a high-spin state with ferromagnetic ordering.

DFT calculations for the *c*MOF cluster and the monomer of *c*P were performed using TeraChem[12]. The B3LYP functional[13-15] was used with the LACVP* basis set, which consists of 6-31G* for elements ranging from H to Ar, and employs the LANL2DZ effective core potential for heavier atoms[16]. To obtain the *c*MOF cluster, one metal node and two linker molecules were extracted from the optimized full MOF structure. Subsequently, hydrogen atoms were added to the truncated bonds, and the positions of the hydrogen atoms were optimized. For calculating the binding energy of the gas molecules, we optimized the positions of the gas molecules while keeping the atomic positions of the *c*MOF cluster fixed. The L-BFGS algorithm was utilized for geometry optimizations via the translation-rotation-internal coordinate optimizer[17,18]. We considered three different metal oxidation states, +0, +1, and +2, each in high spin state. The oxidation state of the metal atom was adjusted by varying the number of hydrogen atoms on the metal-coordinating oxygen atoms, rather than by altering the system's total charge, to neutralize the metal node **(Supplementary Fig. 25c-e)**. This approach was taken because excess negative charge around the metal atom would ionize $NO_2$ to $NO_2^-$ and simply repel away the ion during the structure optimization. Overall charge neutrality was maintained, except for the hole-excess and electron-excess *c*MOF systems which were simulated by removing or adding one electron to the *c*MOF cluster, respectively. To model adsorption, we assumed that $H_2S$ and $NH_3$ adsorbed via the sulfur and nitrogen atoms, respectively, while both N-binding and O-binding were considered as possible adsorbate orientations for $NO_2$ (**Supplementary Fig. 25f-i**). Likewise, the structure of the *c*P monomer was optimized, and the binding energies of the gas molecules were determined by optimizing their positions while fixing the atomic coordinates of the monomer. A negative binding energy indicates favorable binding.

***$H_2S$, $NH_3$, and $NO_2$ selectivity***

We compared the binding energies of $H_2S$, $NH_3$, and $NO_2$ to investigate the source of $NO_2$ selectivity. Across all MOFs of all oxidation states, $NO_2$ generally exhibits stronger binding than $H_2S$ or $NH_3$ (**Supplementary Fig. 26**). Out of 36 total cases, only five exceptions to this trend were observed, specifically in cases involving $Ni_3(HHTP)_2$ and $Cu_3(HHTP)_2$. Furthermore, the binding energy of $NO_2$, considering both N-binding and O-binding, ranges from -3.3 to -0.2 eV, whereas the binding energies of $H_2S$ and $NH_3$ never exceed -1.0 eV. The notable disparity in the range of binding energies suggests that the primary interaction between the metal node and $H_2S$ or $NH_3$ is largely governed by van der Waals interactions, unlike $NO_2$ binding which is enhanced by chemisorption.

***Gas molecule adsorption on cP***

We assessed the binding energies of gas molecules with the *c*P to determine whether the primary adsorption site is the *c*P or *c*MOF. We modeled ProDOT connected to BTD as a monomer and considered eight potential adsorption sites that were neither carbon nor hydrogen atoms. The binding energies of all three gas molecules fell within a range of -0.20 to -0.05 eV, suggesting that the interactions between the *c*P and the gas molecules are primarily weak van der Waals interactions (**Supplementary Fig. 21**). The gas molecules exhibited stronger binding with the *c*MOF than with the *c*P, indicating that the primary adsorption site is likely the metal node of the *c*MOF.

## 3) Supplementary Figures and Tables

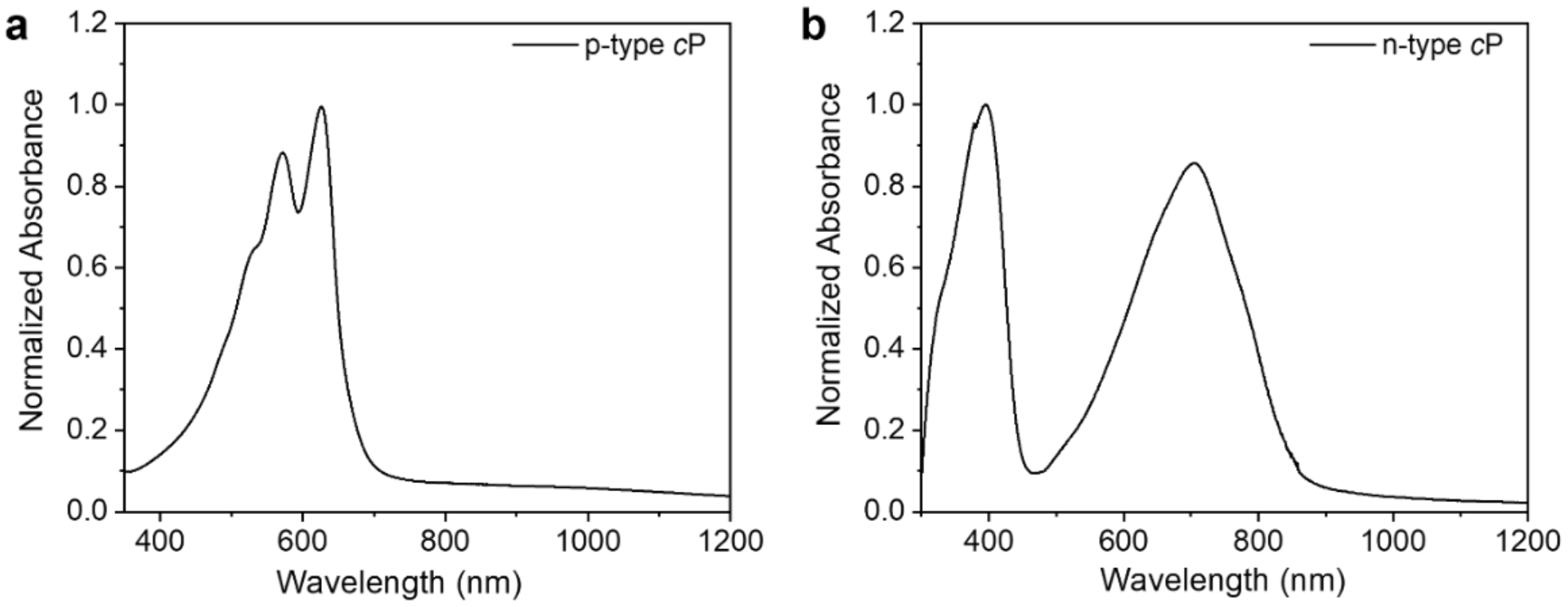

**Supplementary Fig. 1** | UV-vis spectra of p-type *c*P and n-type *c*P.

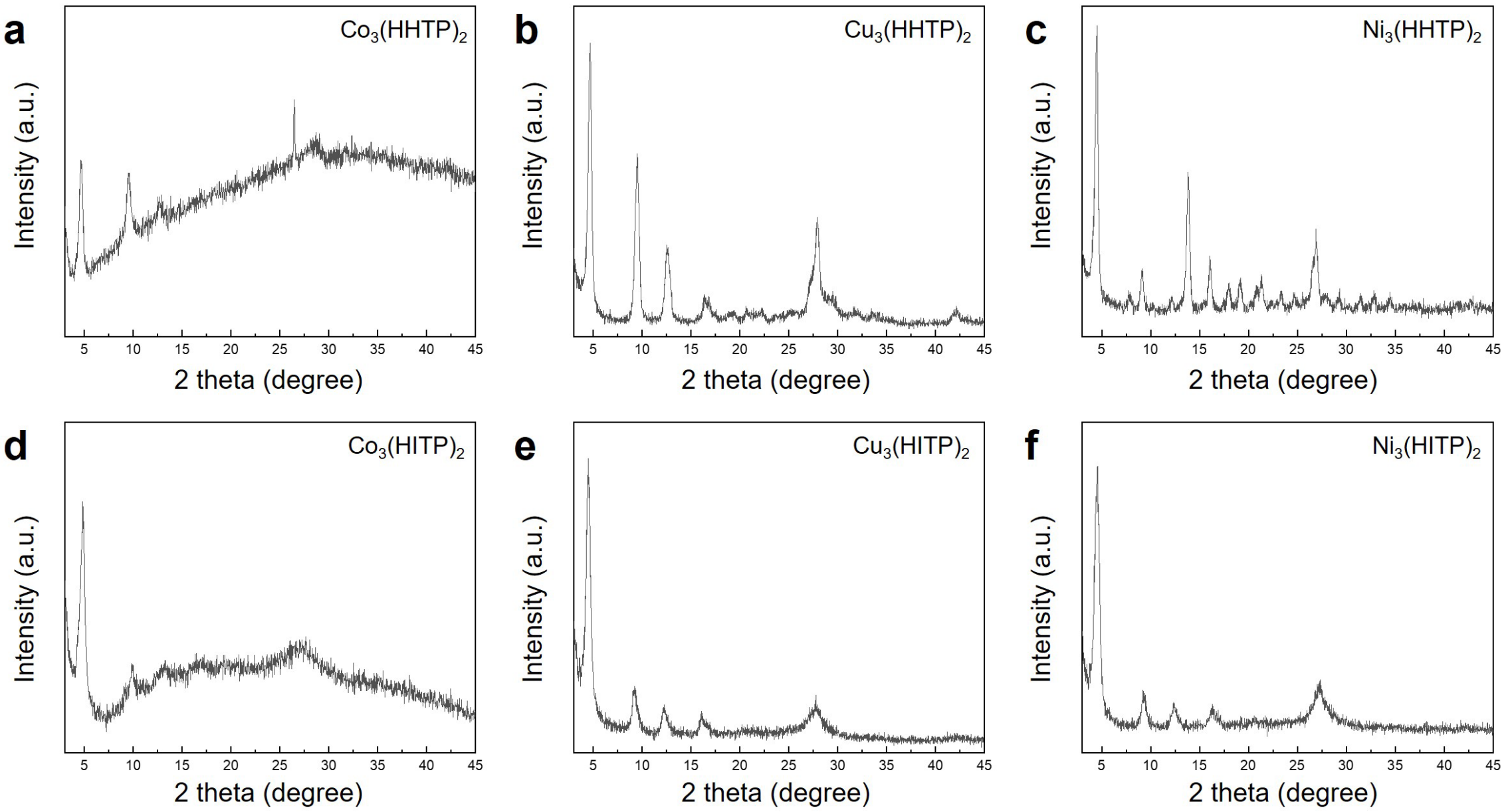

**Supplementary Fig. 2** | XRD analysis of **a,** $Co_3(HHTP)_2$, **b,** $Cu_3(HHTP)_2$, **c,** $Ni_3(HHTP)_2$, **d,** $Co_3(HITP)_2$, **e,** $Cu_3(HITP)_2$, and **f,** $Ni_3(HITP)_2$.

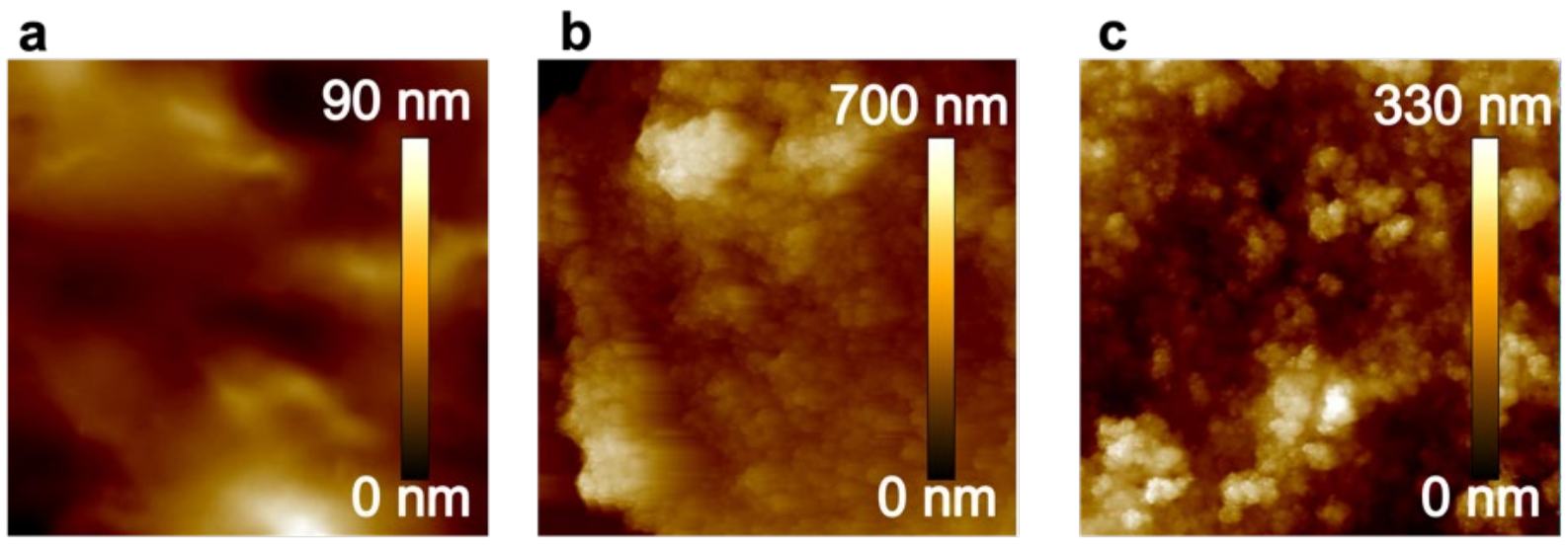

**Supplementary Fig. 3** | Atomic force microscopy (AFM) images were obtained for a 5 μm x 5 μm area and corresponding film roughness parameter (Rq) were determined as 22.7 nm, 195 nm, and 97.5 nm for **a**, *c*P, **b**, *c*MOF, and **c**, *c*P@*c*MOF, respectively.

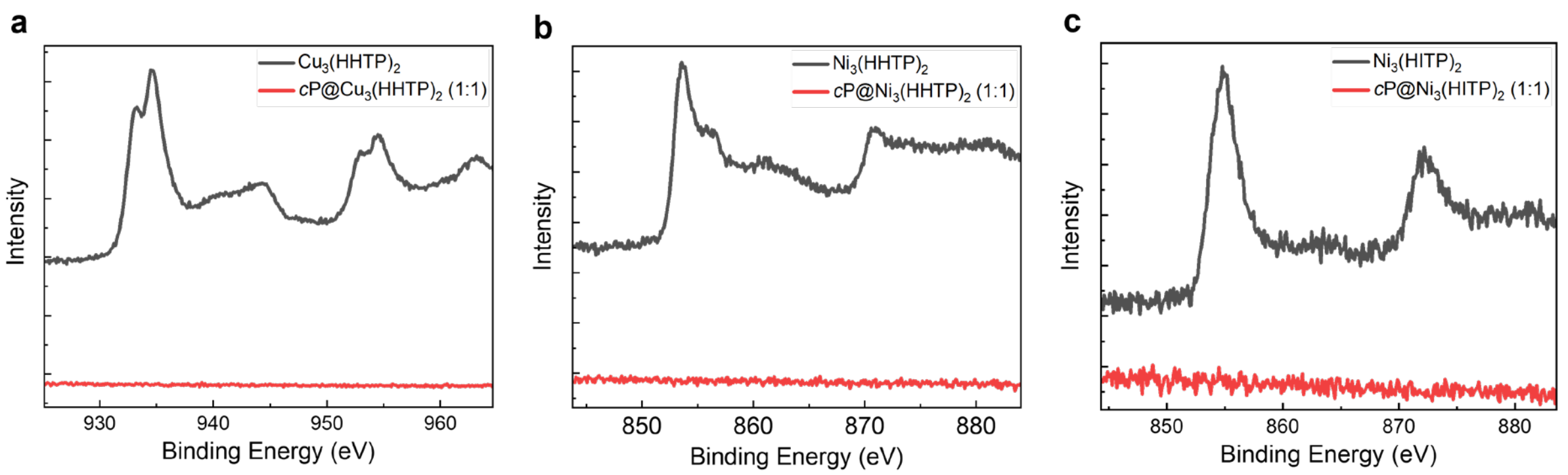

**Supplementary Fig. 4 |** High-resolution XPS spectra of pristine *c*MOFs and *c*P@*c*MOFs hybrids (1:1 w/w% if not mentioned otherwise) using **a**, $Cu_3(HHTP)_2$, **b**, $Ni_3(HHTP)_2$ and **c,** $Ni_3(HITP)_2$-based *c*MOFs, respectively.

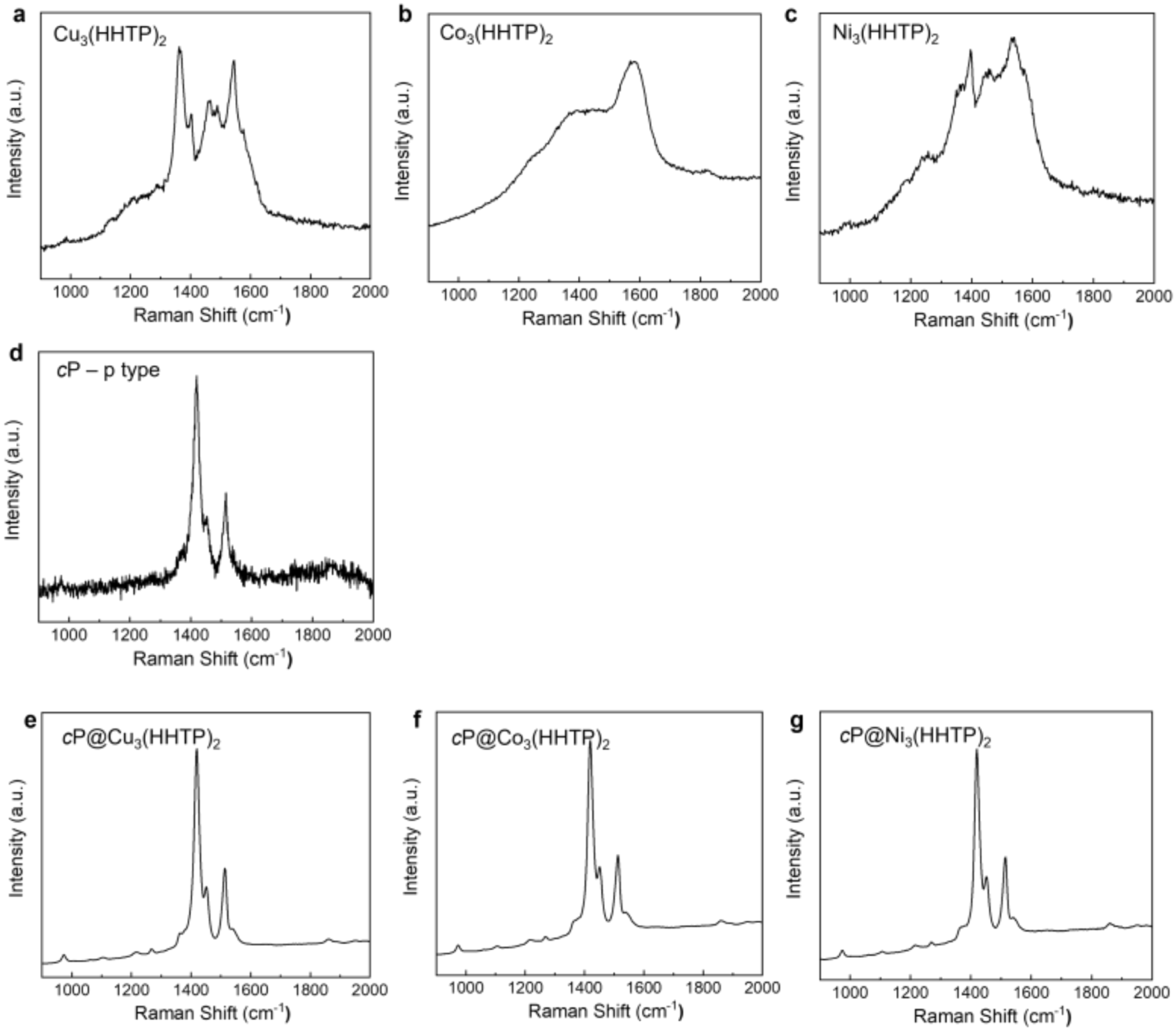

**Supplementary Fig. 5 |** Raman spectra of **a-c**, pristine *c*MOFs, **d**, pristine *c*P, and **e-g**, corresponding hybrid *c*P@*c*MOF (1:1).

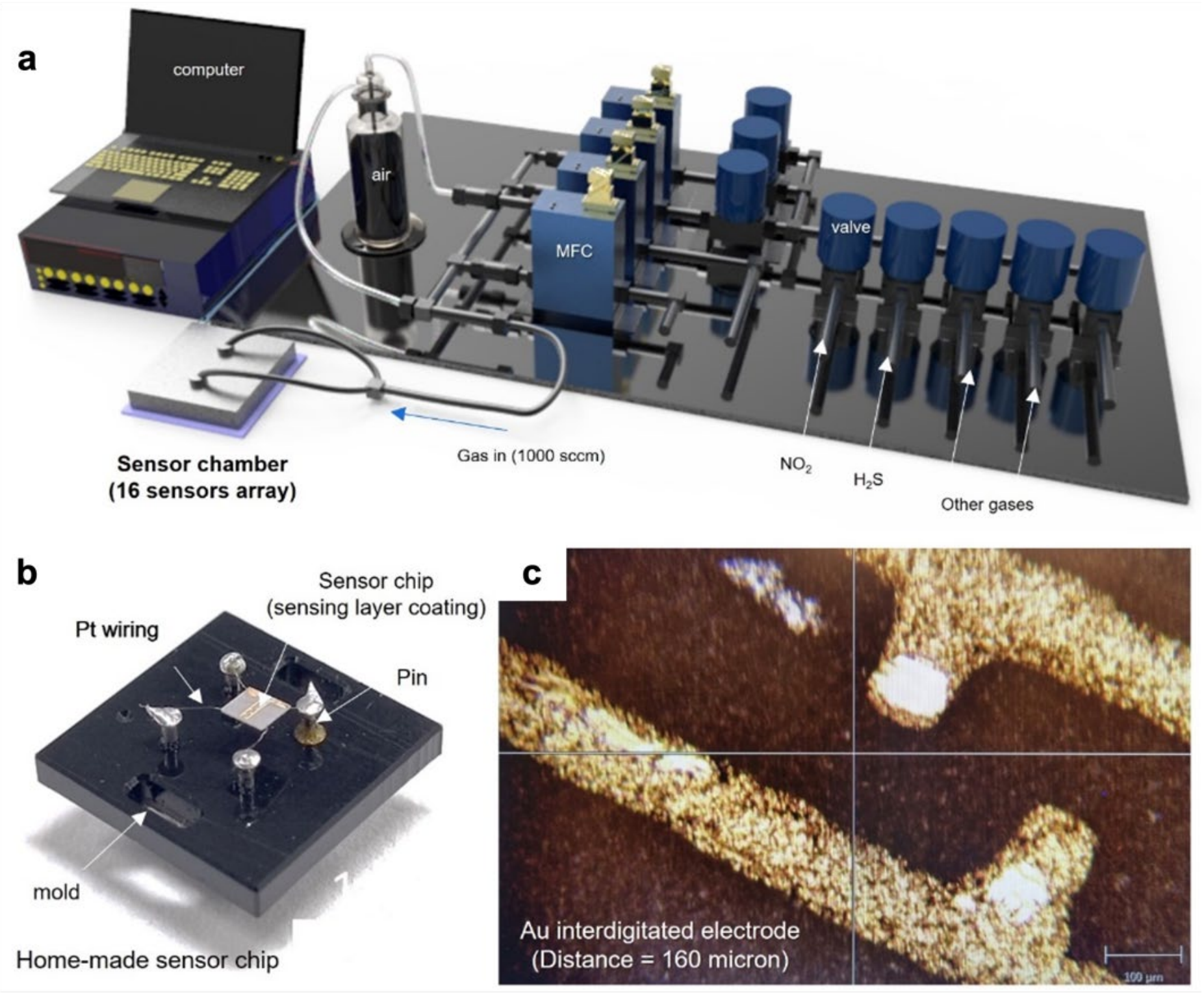

**Supplementary Fig. 6 | a,** Schematic illustrations of the lab-made gas sensing measurement systems. Optical images of the **b,** sensor chip and **c,** sensor electrode.

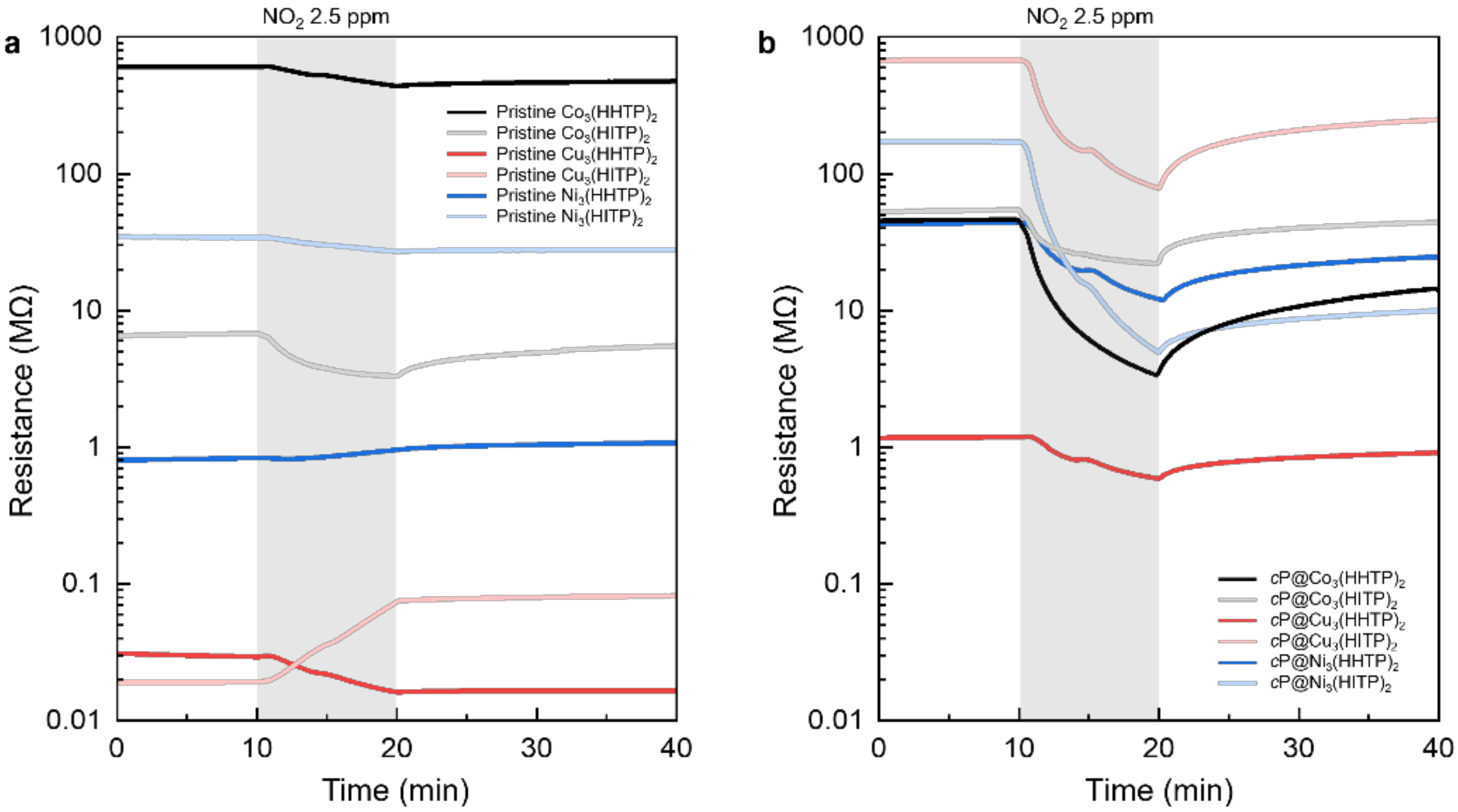

**Supplementary Fig. 7 |** Resistance transitions of **a,** pristine *c*MOFs and **b,** *c*P@*c*MOFs (1:1) toward 2.5 ppm $NO_2$ gas.

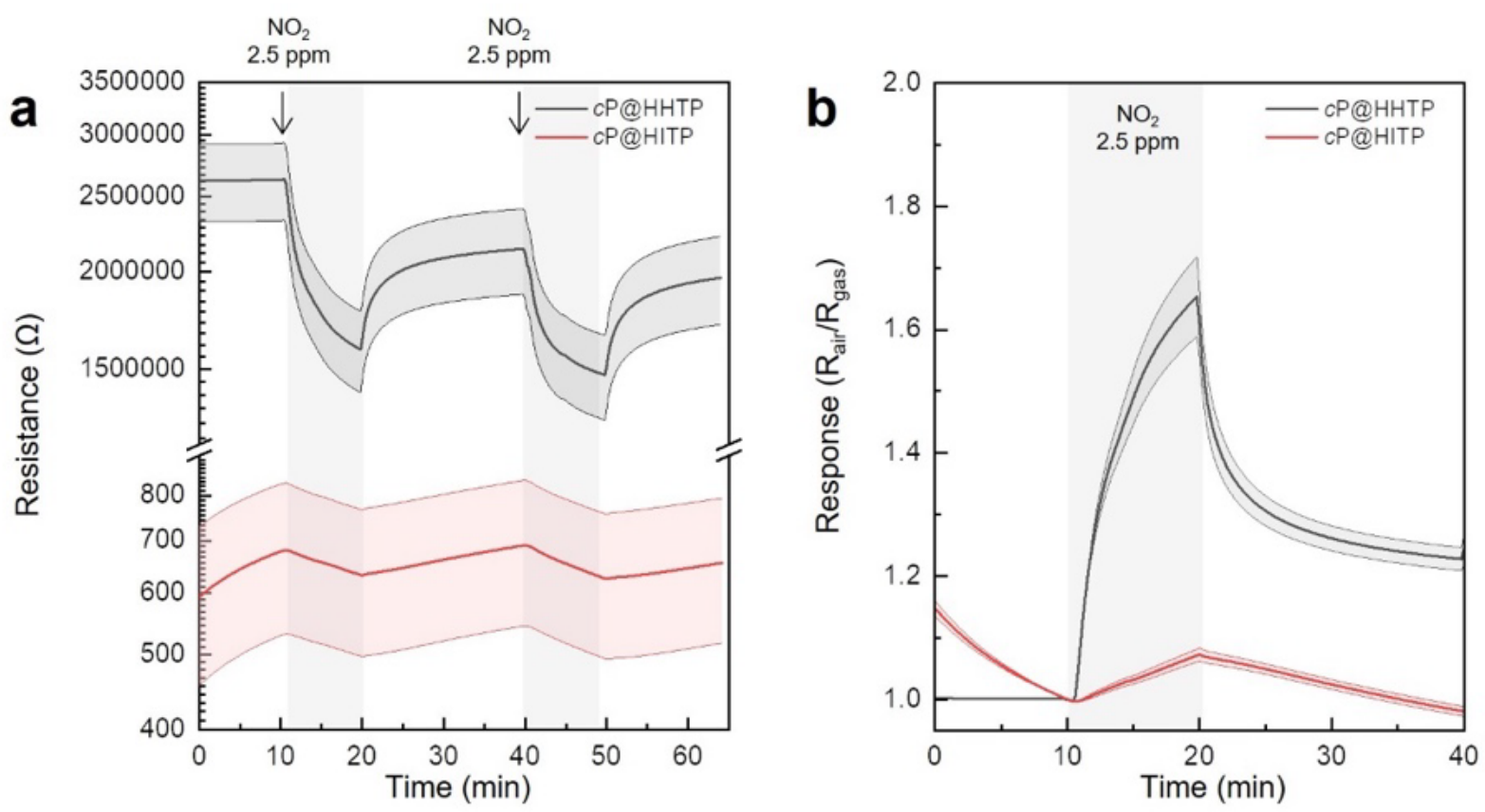

**Supplementary Fig. 8 | a,** Dynamic resistance transitions and **b,** the corresponding response graphs of *c*P@HHTP and *c*P@HITP upon exposure to $NO_2$ gas (N = 3).

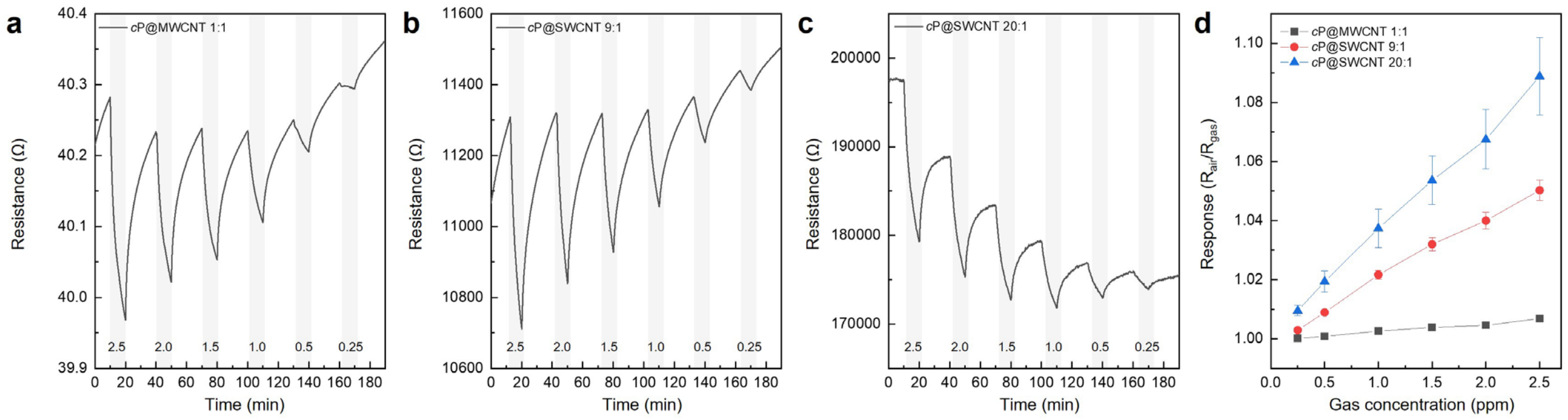

**Supplementary Fig. 9 |** Dynamic resistance transitions of **a,** *c*P@MWCNT 1:1 w/w %, **b,** *c*P@SWCNT 9:1 w/w %, and **c,** *c*P@SWCNT 20:1 w/w % upon exposure to 2.5-0.25 ppm $NO_2$ gas. **d,** The corresponding response graph (N = 3).

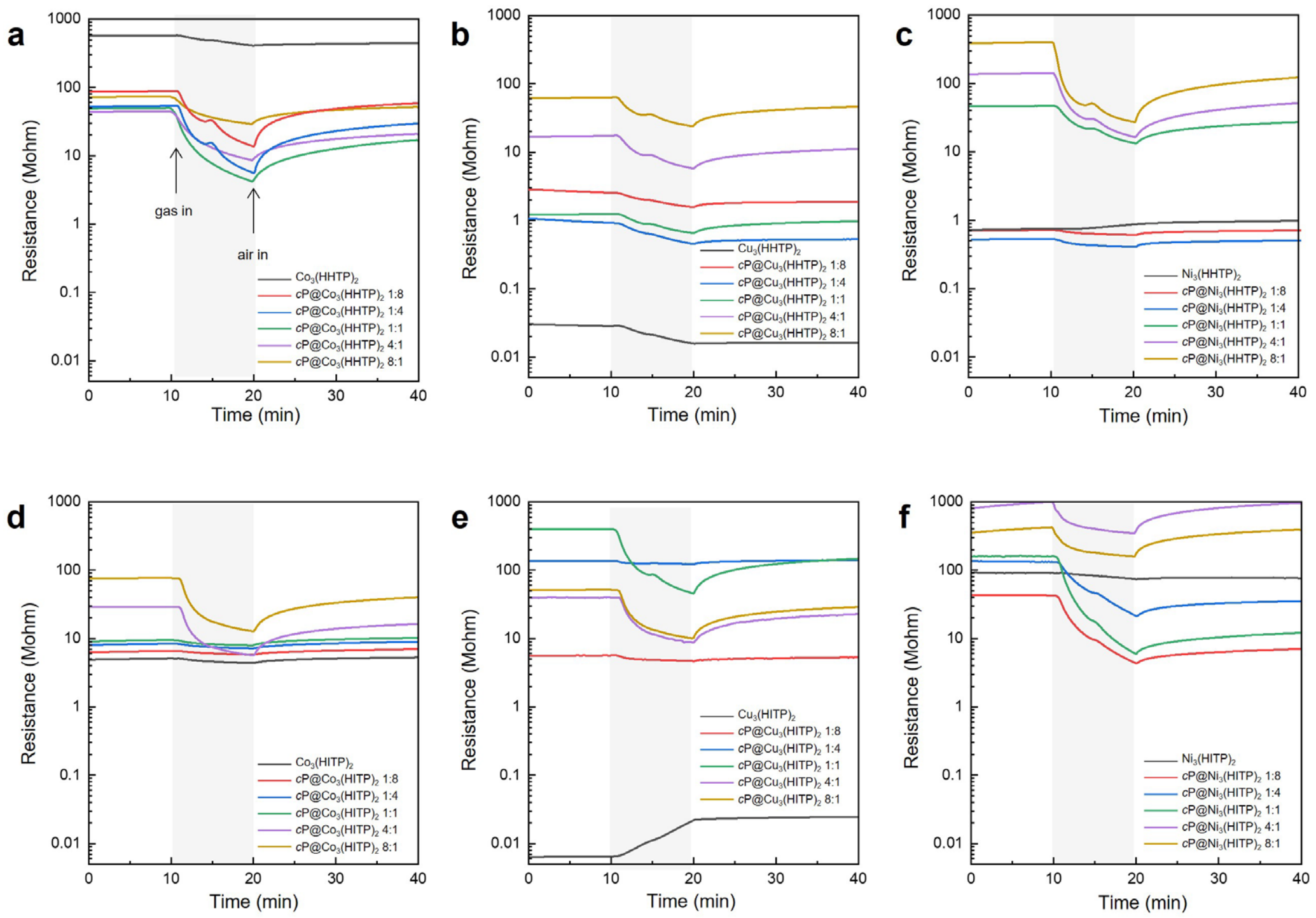

**Supplementary Fig. 10 |** Dynamic resistance transitions of pristine *c*MOFs and *c*P@*c*MOFs toward 2.5 ppm $NO_2$ gas. **a,** $Co_3(HHTP)_2$, **b,** $Cu_3(HHTP)_2$, **c,** $Ni_3(HHTP)_2$, **d,** $Co_3(HITP)_2$, **e,** $Cu_3(HITP)_2$, and **f,** $Ni_3(HITP)_2$-based sensors with various ratios between *c*P and *c*MOFs.

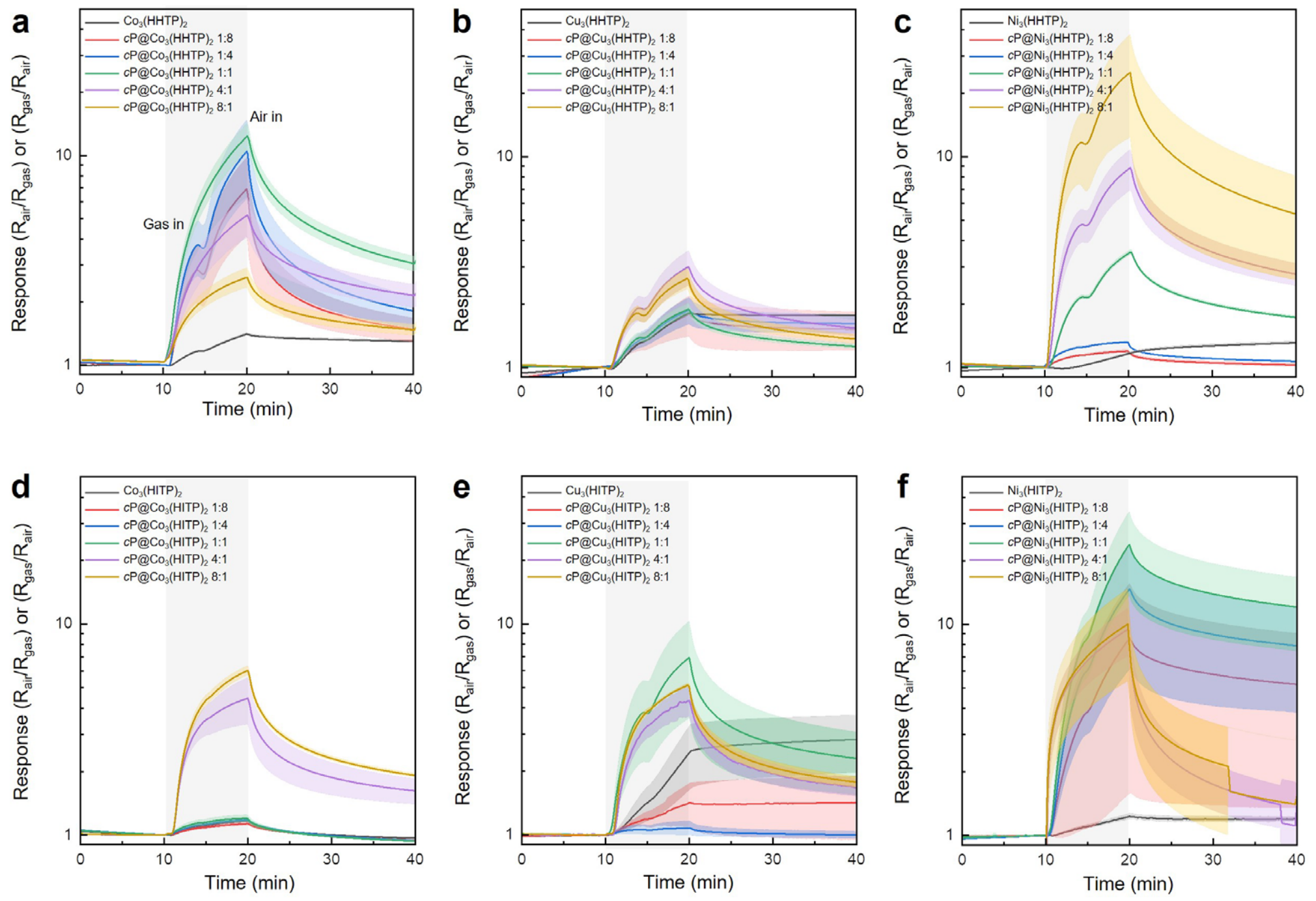

**Supplementary Fig. 11 |** Response graphs of pristine *c*MOFs and *c*P@*c*MOFs. **a,** $Co_3(HHTP)_2$, **b,** $Cu_3(HHTP)_2$, **c,** $Ni_3(HHTP)_2$, **d,** $Co_3(HITP)_2$, **e,** $Cu_3(HITP)_2$, and **f,** $Ni_3(HITP)_2$-based sensors with various ratios between *c*P and *c*MOFs ($N \geq 3$).

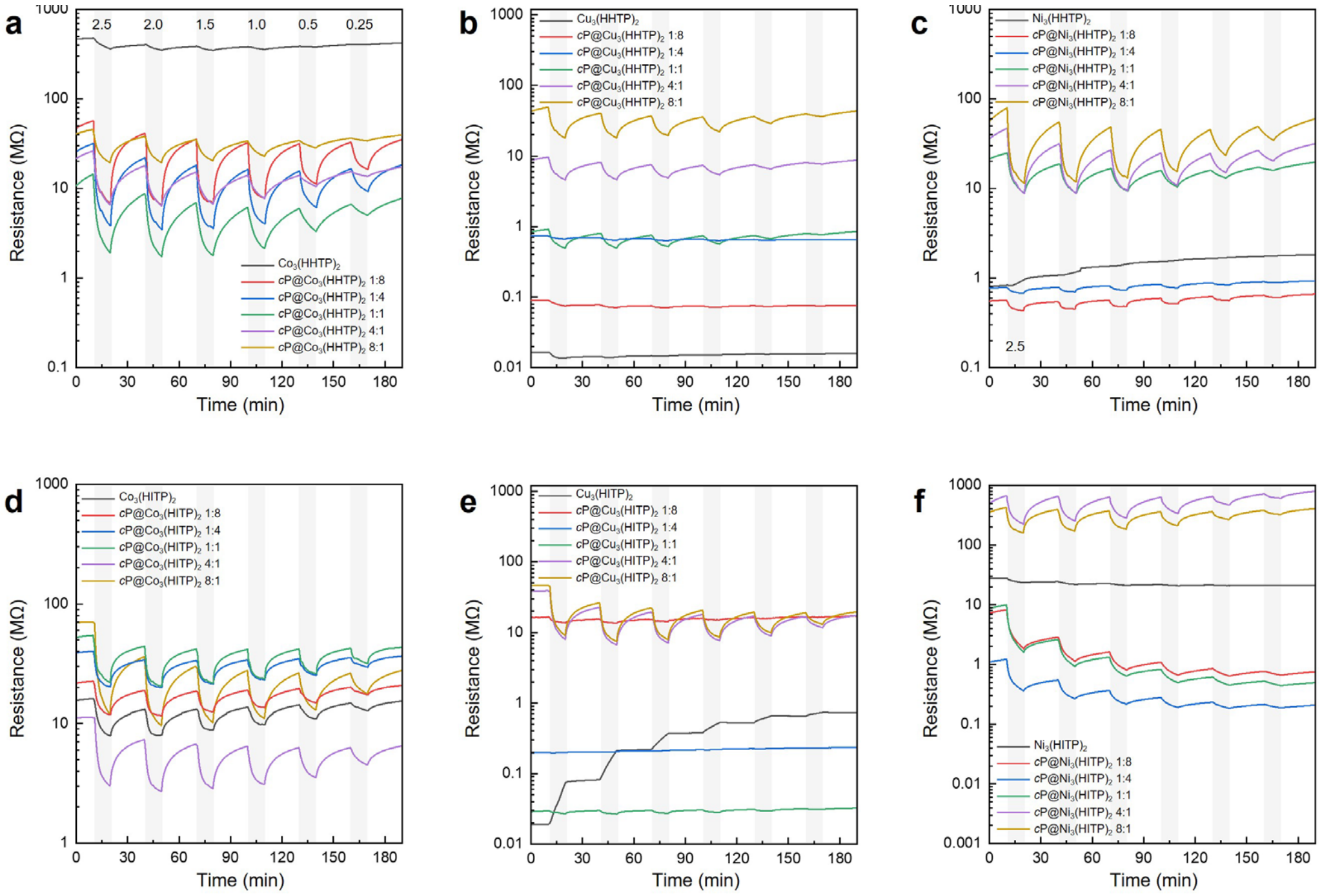

**Supplementary Fig. 12 |** Dynamic resistance transitions of pristine *c*MOFs and *c*P@*c*MOFs toward 2.5-0.25 ppm $NO_2$ gas. **a,** $Co_3(HHTP)_2$, **b,** $Cu_3(HHTP)_2$, **c,** $Ni_3(HHTP)_2$, **d,** $Co_3(HITP)_2$, **e,** $Cu_3(HITP)_2$, and **f,** $Ni_3(HITP)_2$-based sensors with various ratios between *c*P and *c*MOFs.

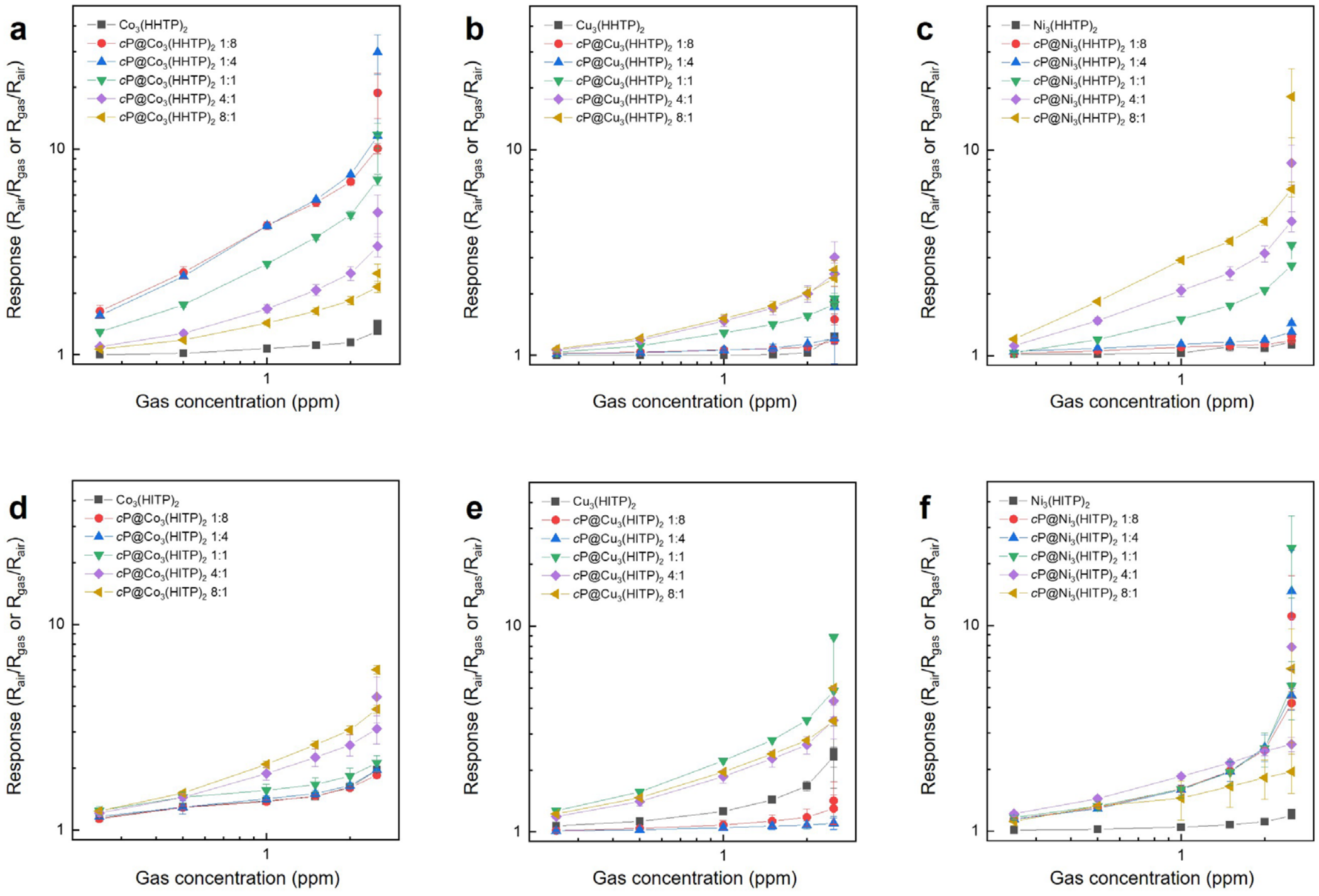

**Supplementary Fig. 13 |** Response graphs of pristine *c*MOFs and *c*P@*c*MOFs toward 2.5-0.25 ppm $NO_2$ gas. **a,** $Co_3(HHTP)_2$, **b,** $Cu_3(HHTP)_2$, **c,** $Ni_3(HHTP)_2$, **d,** $Co_3(HITP)_2$, **e,** $Cu_3(HITP)_2$, and **f,** $Ni_3(HITP)_2$-based sensors with various ratios between *c*P and *c*MOFs ($N \geq 3$).

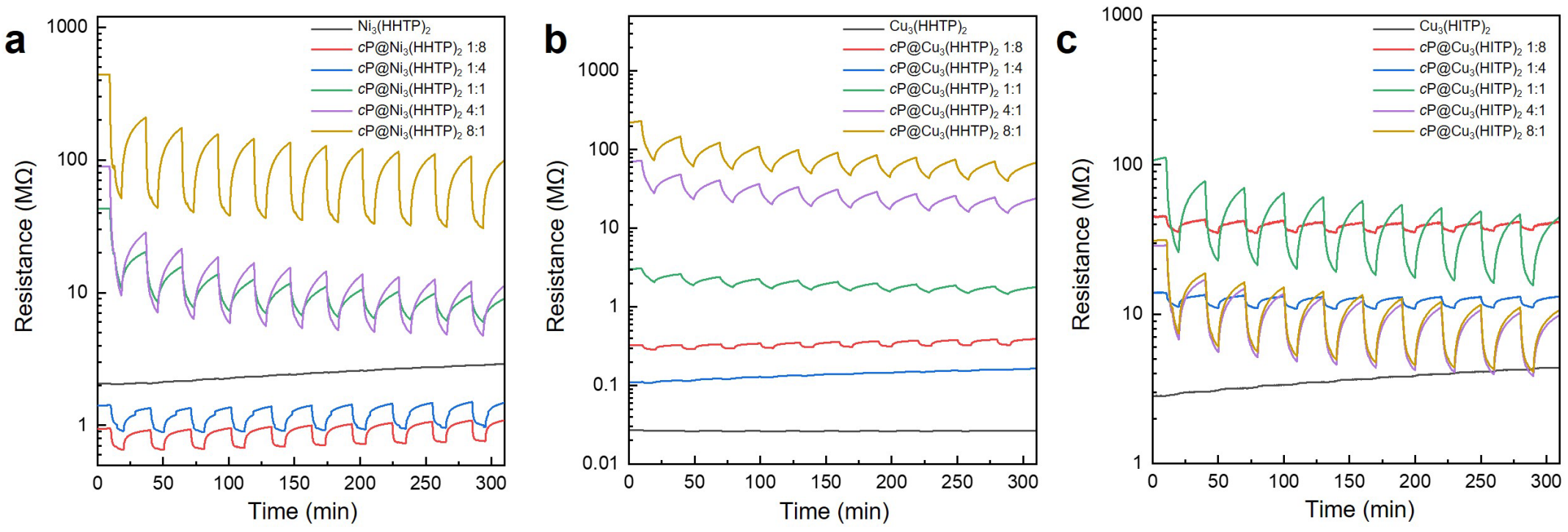

**Supplementary Fig. 14 |** Cyclic sensing tests of **a,** $Ni_3(HHTP)_2$, **b,** $Cu_3(HHTP)_2$, and **c,** $Cu_3(HITP)_2$-based sensors with various ratios between *c*P and *c*MOFs.

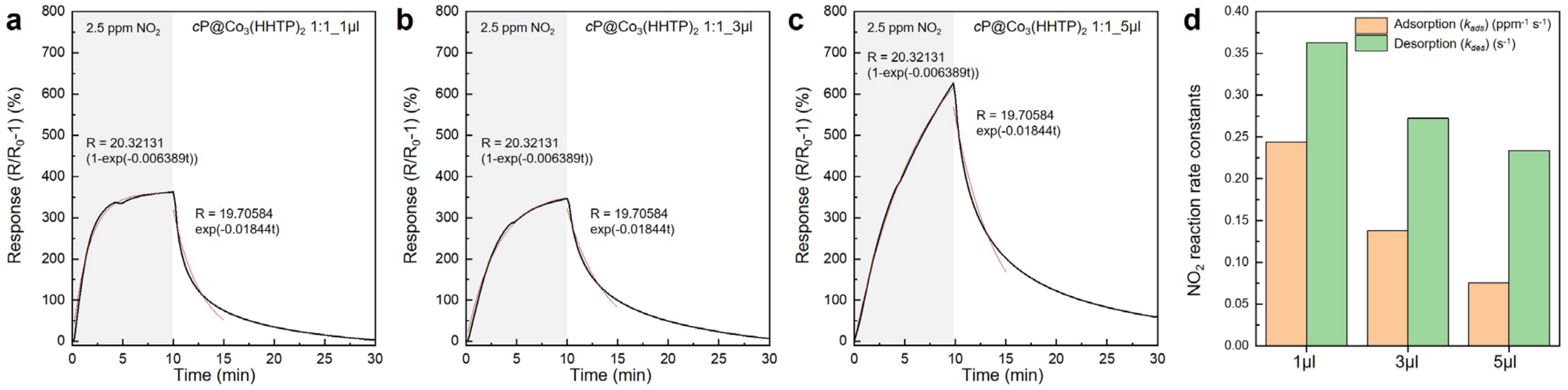

**Supplementary Fig. 15 |** Response and recovery fitting curves and raw response and recovery curves of **a,** *c*P@$Co_3(HHTP)_2$ 1:1_1 µl, **b,** *c*P@$Co_3(HHTP)_2$ 1:1_3 µl, and **c,** *c*P@$Co_3(HHTP)_2$ 1:1_5 µl sensors. **d,** Corresponding adsorption and desorption rate constants of three different sensors. Sensors with reduced coating amounts (1µl) exhibited faster reaction kinetics due to the extended gas diffusion time associated with thicker layers (coating of 3 µl or 5 µl).

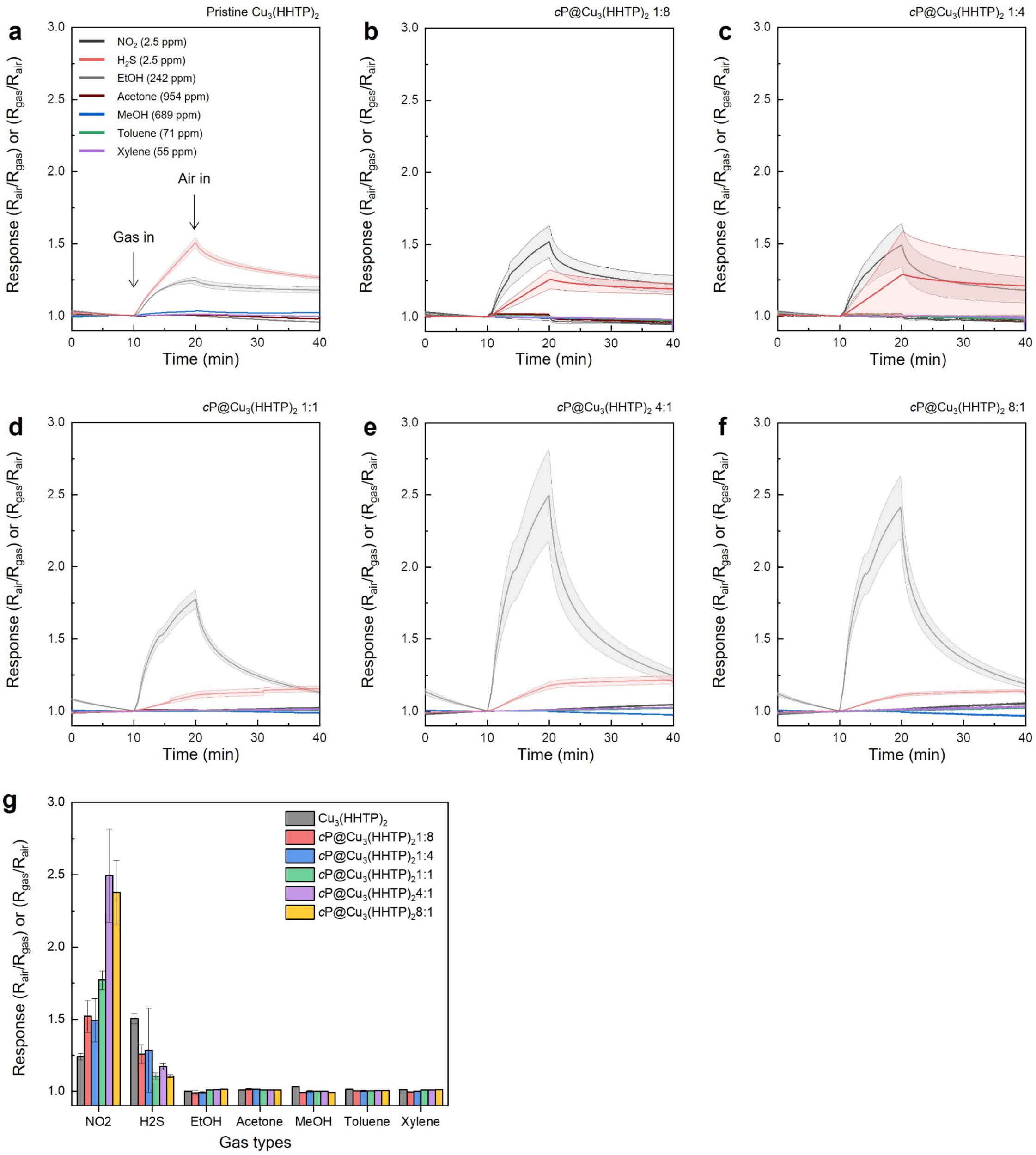

**Supplementary Fig. 16 |** Selectivity of **a,** $Cu_3(HHTP)_2$, **b,** *c*P@$Cu_3(HHTP)_2$ 1:8, **c,** *c*P@$Cu_3(HHTP)_2$ 1:4, **d,** *c*P@$Cu_3(HHTP)_2$ 1:1, **e,** *c*P@$Cu_3(HHTP)_2$ 4:1, and **f,** *c*P@$Cu_3(HHTP)_2$ 8:1. **g,** Overall response graphs ($N \geq 3$).

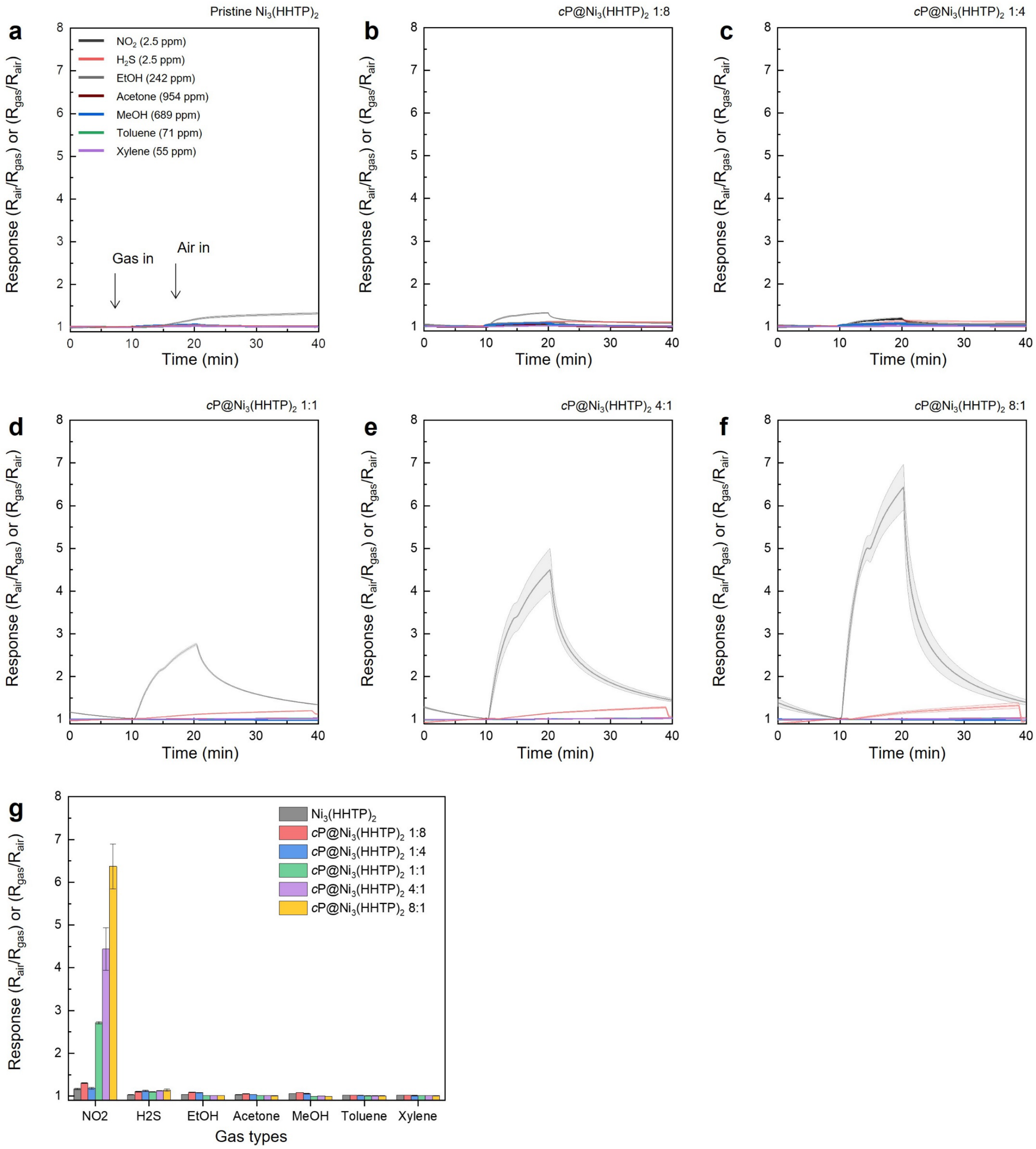

**Supplementary Fig. 17 |** Selectivity of **a,** $Ni_3(HHTP)_2$, **b,** *c*P@$Ni_3(HHTP)_2$ 1:8, **c,** *c*P@$Ni_3(HHTP)_2$ 1:4, **d,** *c*P@$Ni_3(HHTP)_2$ 1:1, **e,** *c*P@$Ni_3(HHTP)_2$ 4:1, and **f,** *c*P@$Ni_3(HHTP)_2$ 8:1. **g,** Overall response graphs ($N \geq 3$).

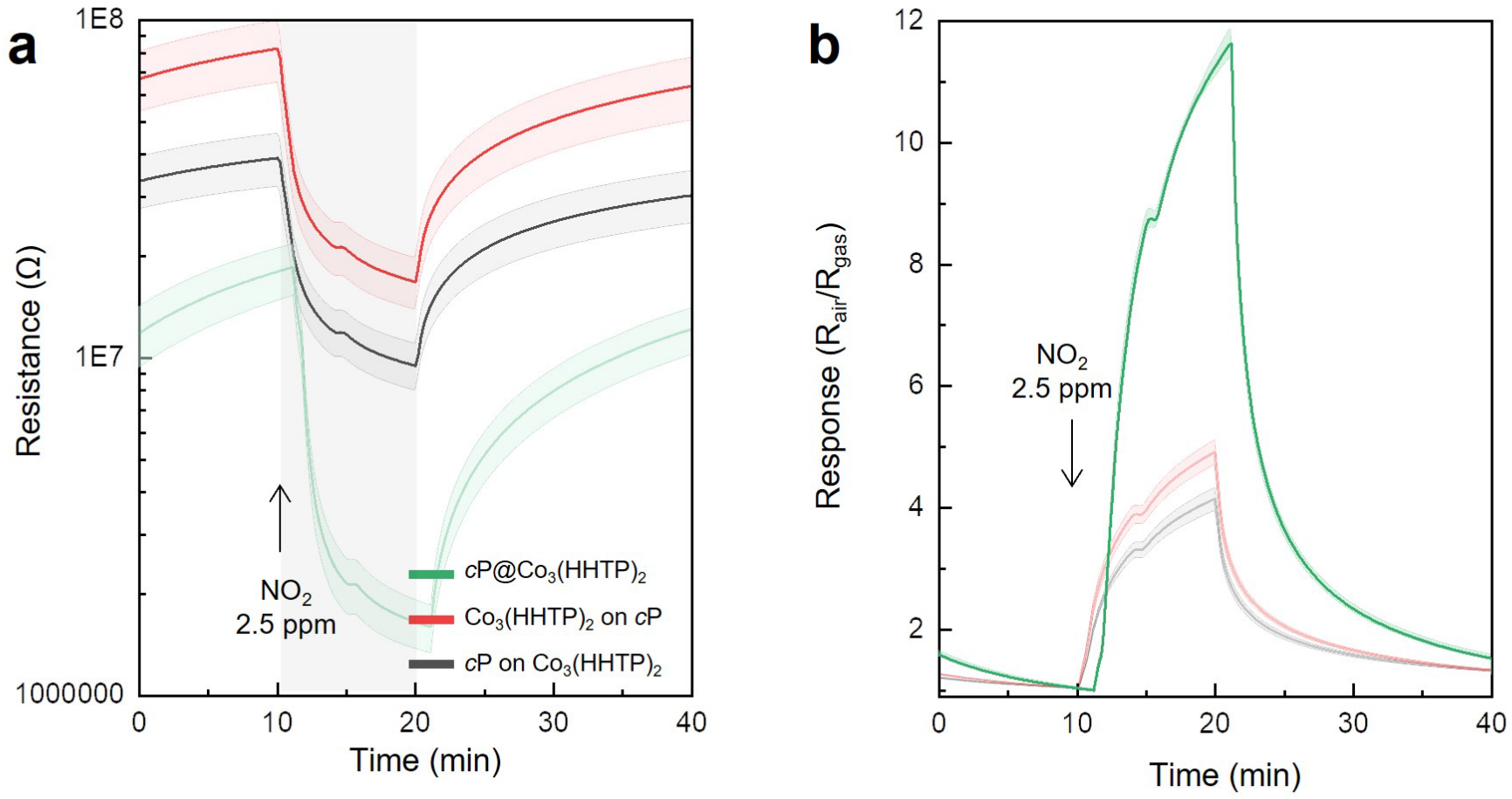

**Supplementary Fig. 18 | a,** Dynamic resistance transitions and **b,** response graphs of *c*P@$Co_3(HHTP)_2$ prepared by three different methods; i) finely mixed *c*P@$Co_3(HHTP)_2$, ii) $Co_3(HHTP)_2$ overlayer on *c*P, and iii) *c*P overlayer on $Co_3(HHTP)_2$ (N = 3).

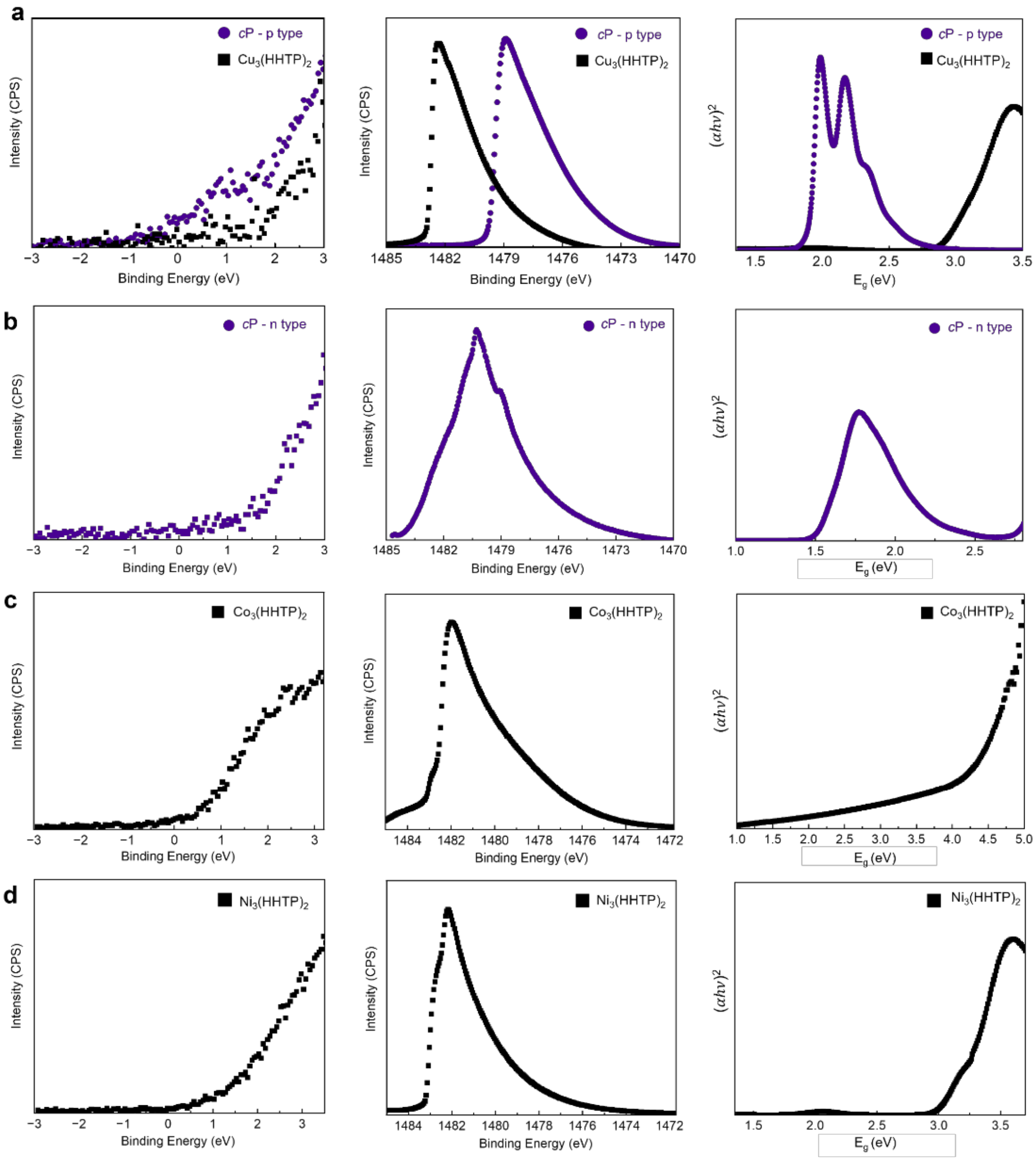

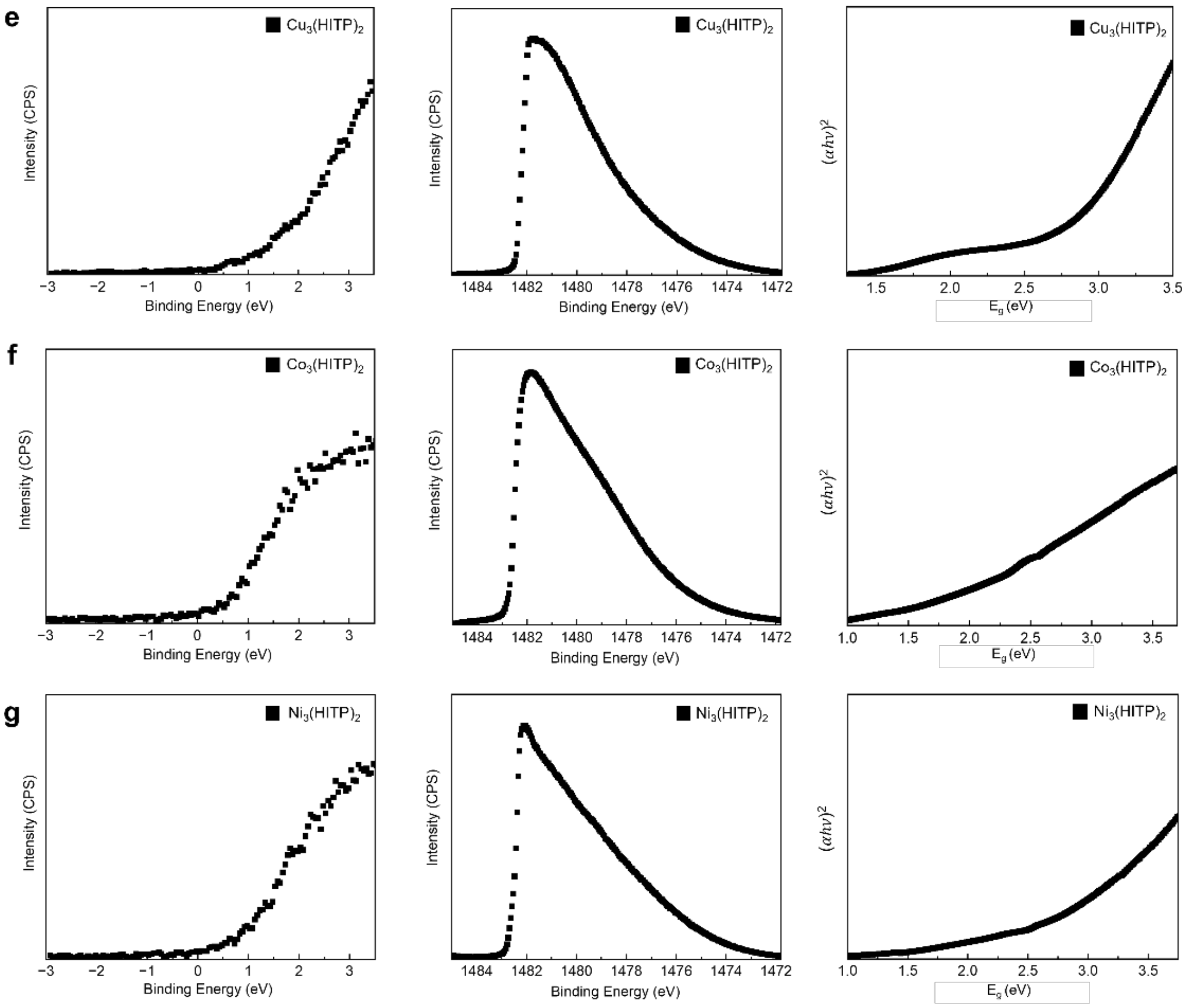

**Supplementary Fig. 19 |** XPS Cutoff curves (left), Fermi curves (middle), and Tauc plots (right), respectively, of *c*MOFs and *c*P. **a,** p-type *c*P and $Cu_3(HHTP)_2$. **b,** n-type *c*P. **c,** $Co_3(HHTP)_2$. **d,** $Ni_3(HHTP)_2$ **e,** $Cu_3(HITP)_2$. **f,** $Co_3(HITP)_2$. **g,** $Ni_3(HITP)_2$.

**Supplementary Table. S1 |** Summary of experimentally measured energy level details.

| | **p-type *c*P** | **n-type *c*P** | **$Cu_3(HHTP)_2$** | **$Ni_3(HHTP)_2$** | **$Co_3(HHTP)_2$** | **$Cu_3(HITP)_2$** | **$Ni_3(HITP)_2$** | **$Co_3(HITP)_2$** |
|---|---|---|---|---|---|---|---|---|
| Photon E | 1486.68 eV | 1486.68 eV | 1486.68 eV | 1486.68 eV | 1486.68 eV | 1486.68 eV | 1486.68 eV | 1486.68 eV |
| Cutoff KE | 11.25 eV | 11.34 eV | 13.77 eV | 13.44 eV | 13.89 eV | 14.24 eV | 14.05 eV | 13.91 eV |
| Fermi KE | 1492 eV | 1490.05 eV | 1496.29 eV | 1495.34 eV | 1495.76 eV | 1496.02 eV | 1496.15 eV | 1496.3 eV |
| Work Function | 6.05 eV | 7.97 eV | 4.15 eV | 4.77 eV | 4.81 eV | 4.89 eV | 4.580 eV | 4.29 eV |
| $E_F$-$E_V$ | 0.31 eV | 1.39 eV | 1.25 eV | 1.35 eV | 0.51 eV | 1.4 eV | 0.92 eV | 0.52 eV |
| $E_g$ | 1.83 eV | 1.53 eV | 2.88 eV | 2.87 eV | 3.9 eV | 2.41 eV | 3.9 eV | 2.54 eV |

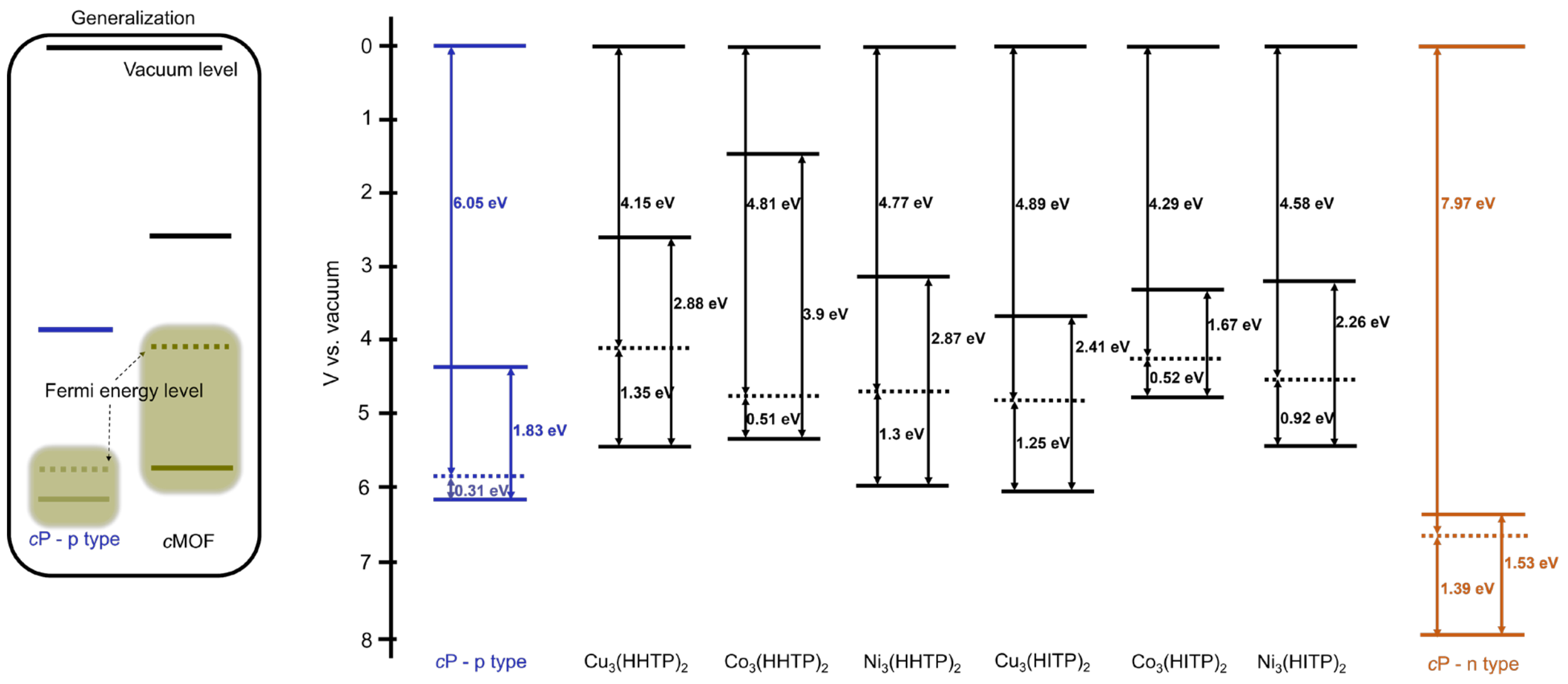

**Supplementary Fig. 20 |** Energy levels of *c*P and *c*MOFs used in the present work.

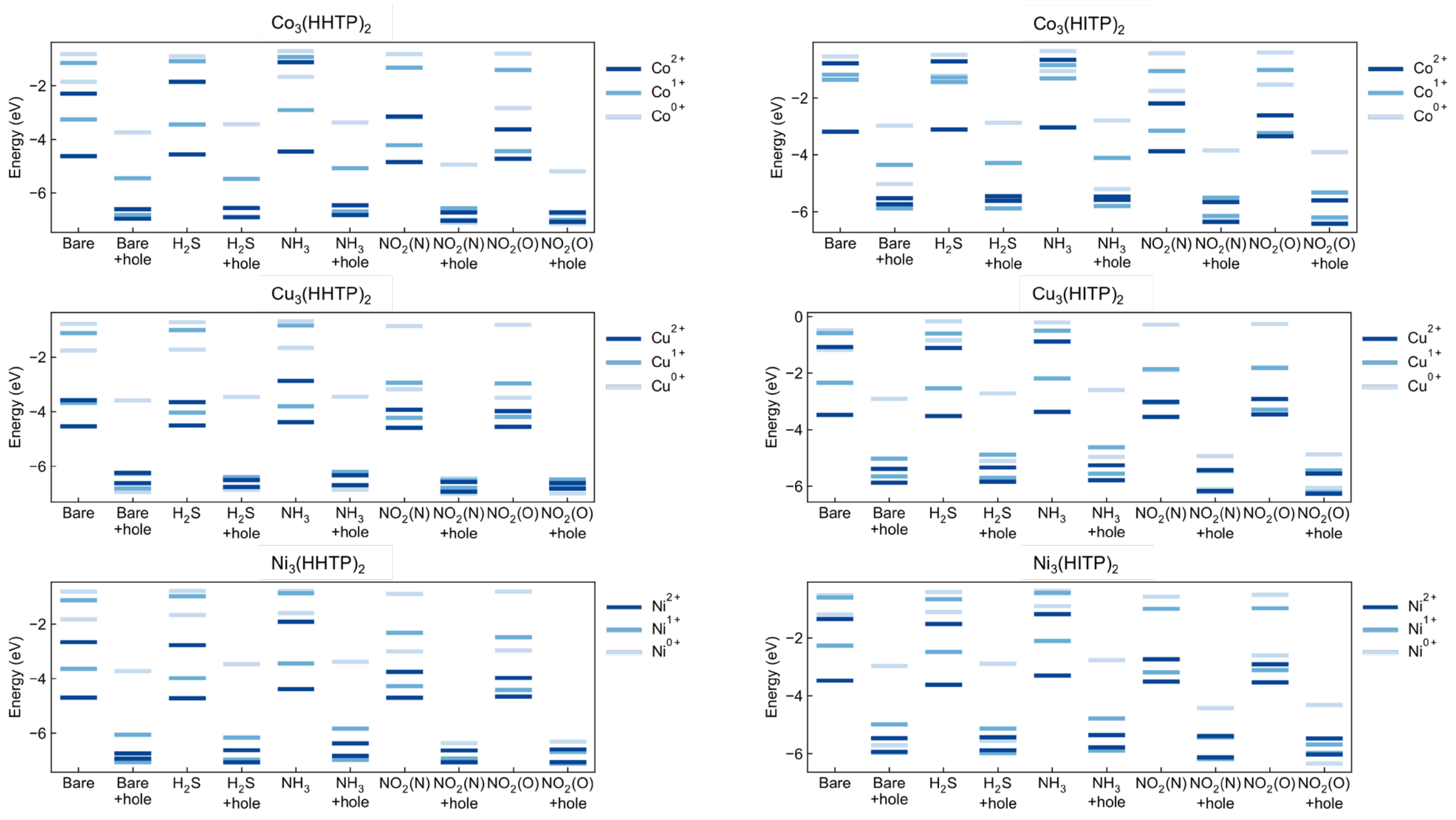

**Supplementary Fig. 21 |** HOMO and LUMO levels of cMOFs both with and without a hole, as well as with and without an adsorbate. All energy levels were calculated using the B3LYP functional and LACVP* basis set. NO2(N) and NO2(O) indicate N-binding and O-binding or NO2, respectively. The line color denotes the oxidation state of the metal.

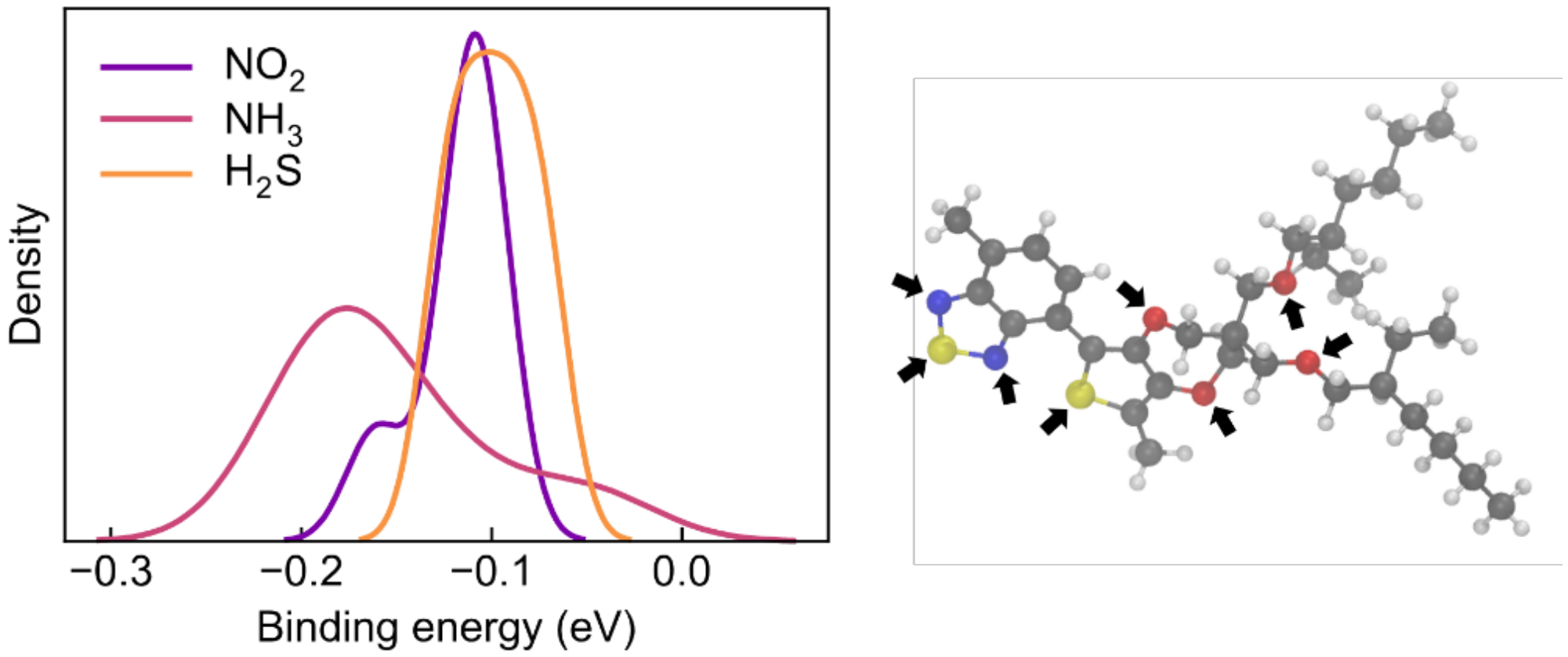

**Supplementary Fig. 22 |** Distribution of binding energies for $NO_2$, $NH_3$, and $H_2S$ to different sites on the monomer of *c*P, as indicated by the arrow. N, O, and S atoms are considered as potential binding sites. The full monomer-adsorbate geometry was optimized using the B3LYP functional and LACVP* basis set. O: Red, N: Blue, H: White, S: Yellow.

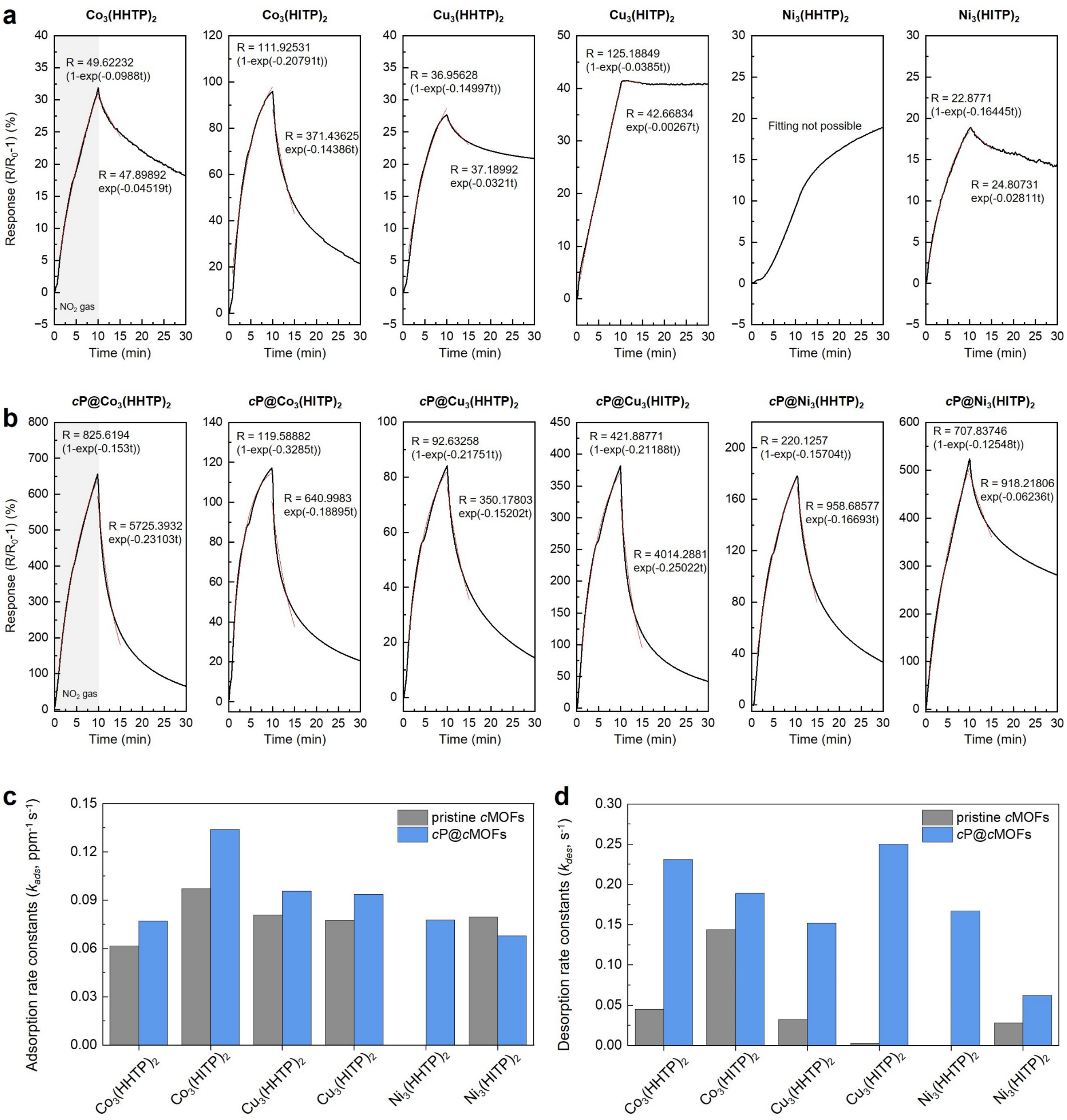

**Supplementary Fig. 23 |** Response and recovery fitting curves of **a,** six different *c*MOFs and **b,** *c*P@*c*MOFs 1:1 composite. The corresponding **c,** adsorption and **d,** desorption rate constants.

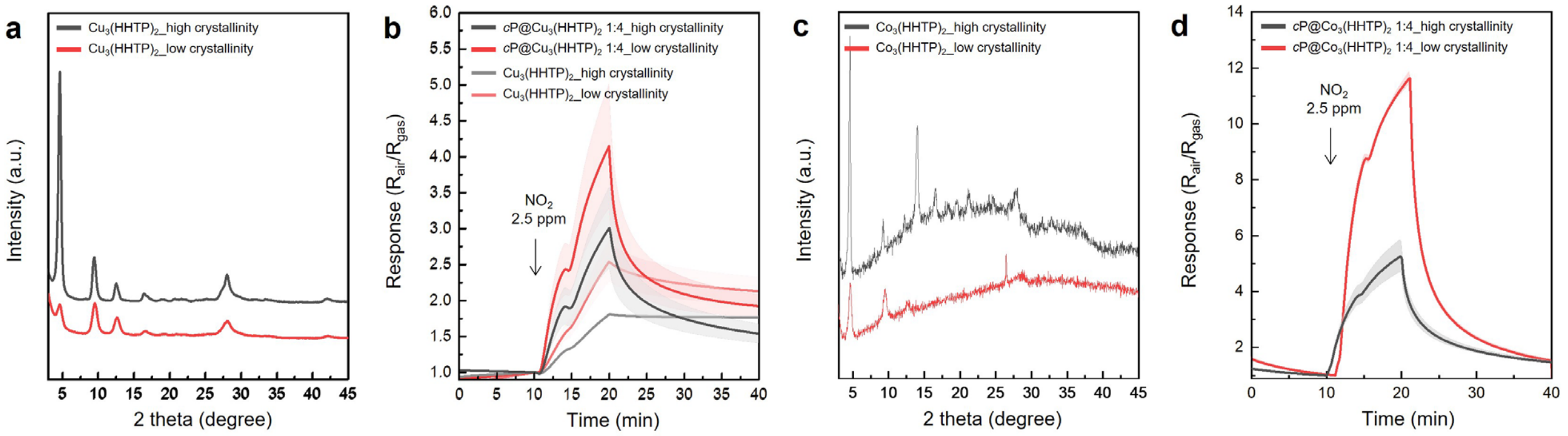

**Supplementary Fig. 24 | a,** XRD analysis of $Cu_3(HHTP)_2$ with high and low crystallinity. **b,** $NO_2$ sensing response of *c*P@$Cu_3(HHTP)_2$ and pristine $Cu_3(HHTP)_2$ using high and low crystallinity of *c*MOFs (N ≥ 3). **c,** XRD analysis of $Co_3(HHTP)_2$ with high and low crystallinity. **d,** $NO_2$ sensing response of *c*P@$Co_3(HHTP)_2$ with high and low crystallinity (N ≥ 3).

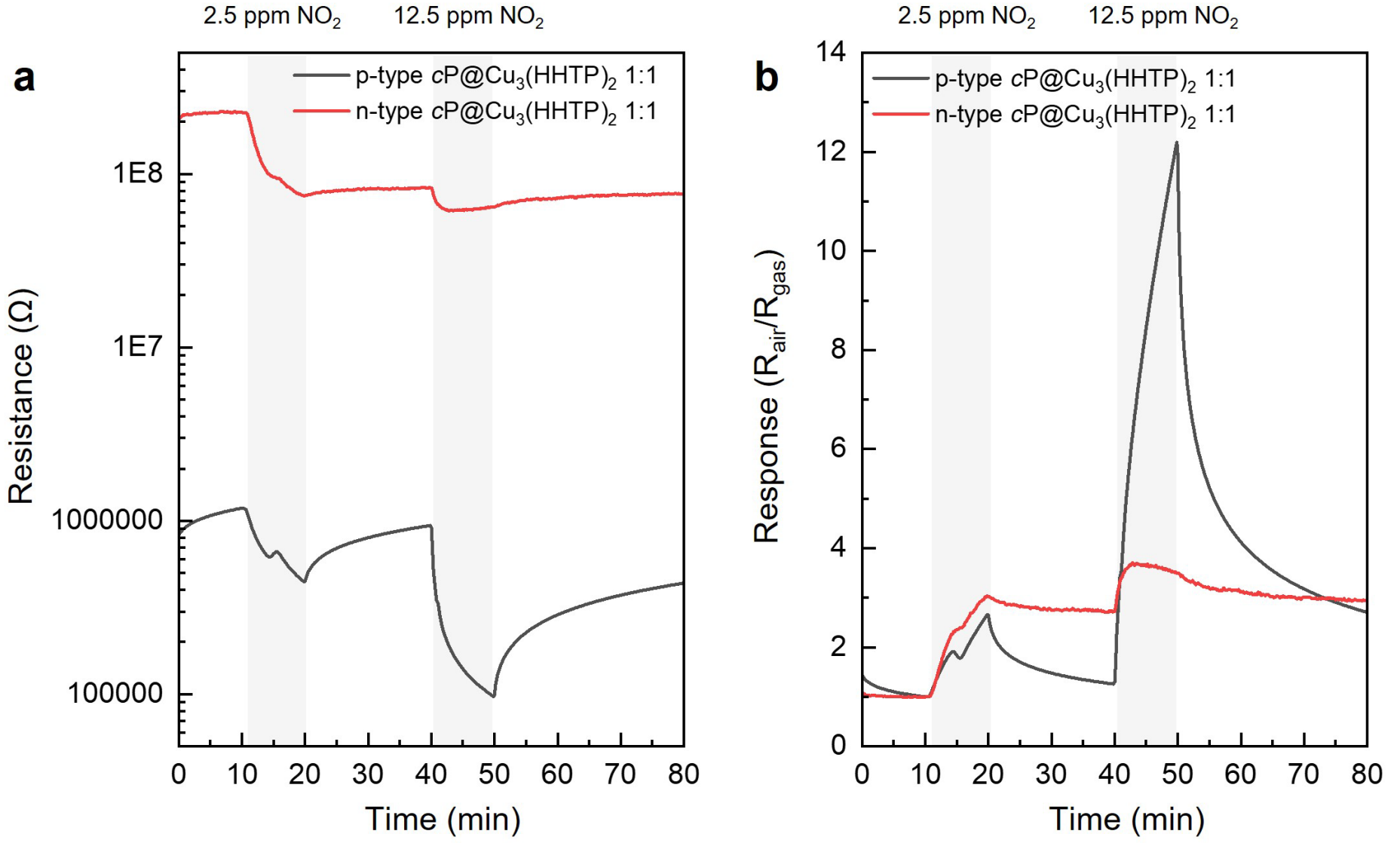

**Supplementary Fig. 25 | a**, Dynamic resistance transitions and **b**, response graphs of p-type *c*P@$Cu_3(HHTP)_2$ 1:1 and n-type *c*P@$Cu_3(HHTP)_2$ 1:1 upon sequential exposure to 2.5 ppm and 12.5 ppm $NO_2$ gas.

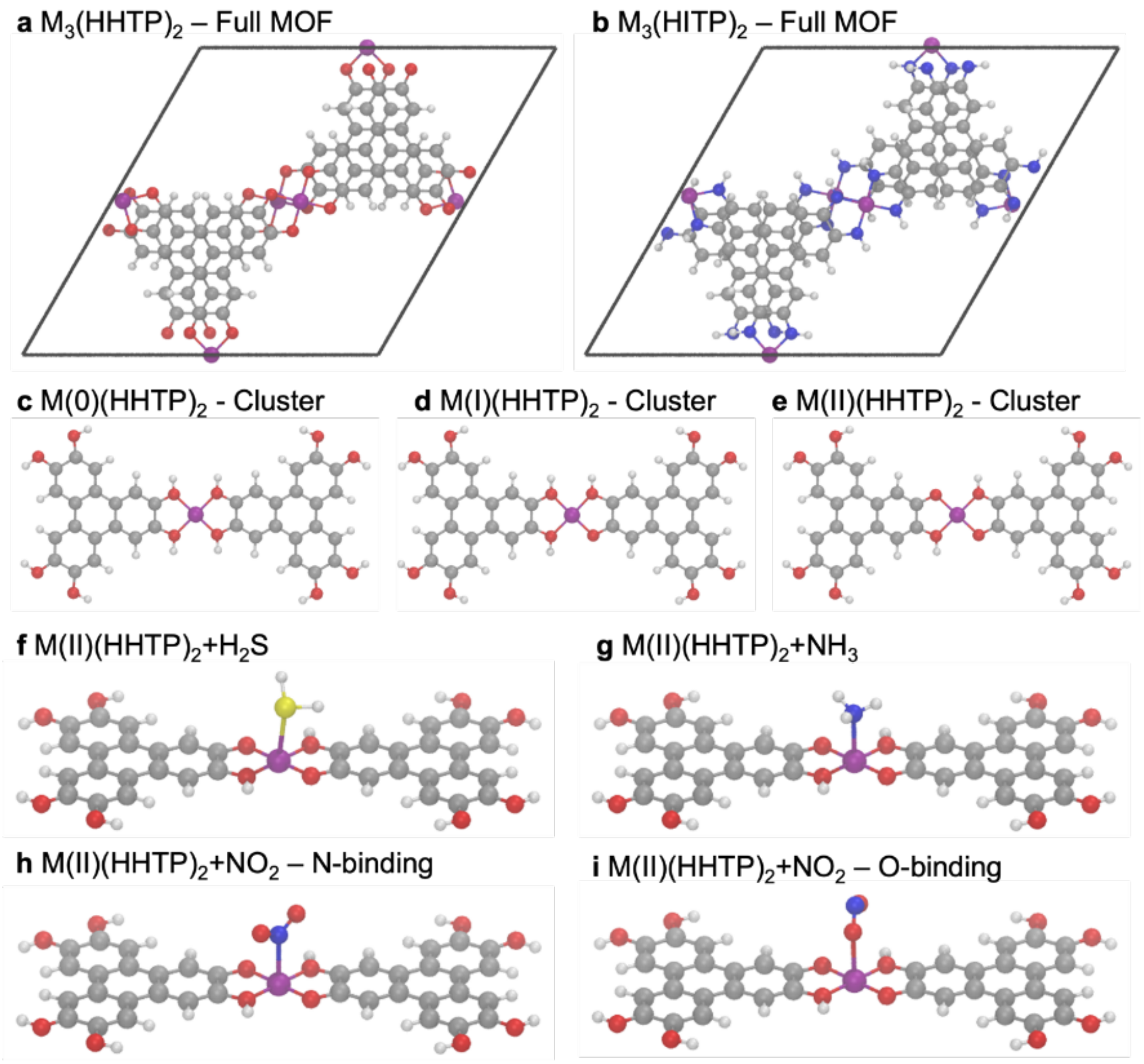

**Supplementary Fig. 26 |** Top view of the complete *c*MOF structures for **a**, $M_3(HHTP)_2$ and **b**, $M_3(HITP)_2$.The black outline indicates the boundaries of the unit cell. Clusters of $M_3(HHTP)_2$ are shown for various oxidation state of the metal: **c**, +0, **d**, +1, and **e**, +2. Note the differing numbers of hydrogen atoms on the metal-coordinating oxygens used to alter the metal oxidation state. Clusters of $M_3(HHTP)_2$ are also illustrated while binding with molecules: **f**, $H_2S$, **g**, $NH_3$, **h**, $NO_2$ with nitrogen atom binding, and **i**, $NO_2$ with oxygen atom binding. M: Purple, C: Grey, O: Red, N: Blue, H: White.

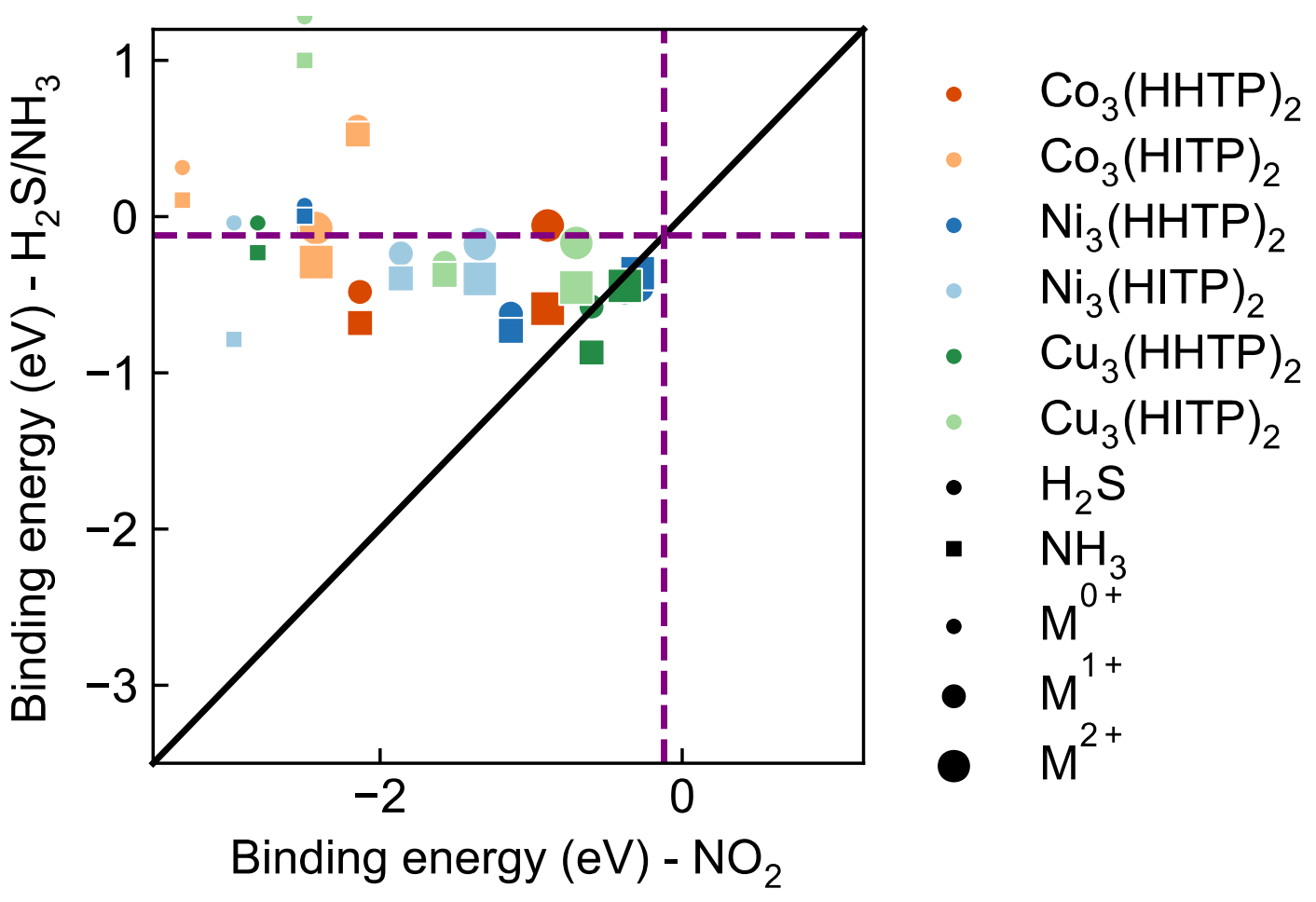

**Supplementary Fig. 27 |** Binding energy of $H_2S$ and $NH_3$ to the MOF compared with that of $NO_2$. The purple dashed line represents the average binding energy between the *c*P monomer and the gas molecules. A black line indicates parity.

## 4) Supplementary References